%% file: aSI.tex
\documentclass[twocolumn]{IEEEtran}
\usepackage{amsmath,amssymb,amsthm,epsfig,color,subfigure,empheq,graphicx,graphics,balance}
\usepackage{enumerate,url,algorithm,algorithmic,wasysym,epstopdf,enumitem}

\input{stylesV2}

\begin{document}
\title{Strategic Investment in Energy Markets:\\
A Multiparametric Programming Approach}

\author{
	Sina Taheri~\IEEEmembership{Student Member,~IEEE},
    Vassilis Kekatos,~\IEEEmembership{Senior Member,~IEEE}, and 
    Harsha Veeramachaneni

    

}

\markboth{IEEE TRANSACTIONS ON POWER SYSTEMS (submitted \today)}{Taheri, Kekatos, and Veermachaneni: Strategic Investment in Energy Markets: A Multiparametric Programming Approach}

\maketitle

\begin{abstract}
An investor has to carefully select the location and size of new generation units it intends to build, since adding capacity in a market affects the profit from units this investor may already own. To capture this closed-loop characteristic, strategic investment (SI) can be posed as a bilevel optimization. By analytically studying a small market, we first show that its objective function can be non-convex and discontinuous. Realizing that existing mixed-integer problem formulations become impractical for larger markets and increasing number of scenarios, this work put forth two SI solvers: a grid search to handle setups where the candidate investment locations are few, and a stochastic gradient descent approach for otherwise. Both solvers leverage the powerful toolbox of multiparametric programming (MPP), each in a unique way. The grid search entails finding the primal/dual solutions for a large number of optimal power flow (OPF) problems, which nonetheless can be efficiently computed several at once thanks to the properties of MPP. The same properties facilitate the rapid calculation of gradients in a mini-batch fashion, thus accelerating the implementation of a stochastic gradient descent search. Tests on the IEEE 118-bus system using real-world data corroborate the advantages of the novel MPP-aided solvers. 
\end{abstract}


\allowdisplaybreaks

\section{Introduction}\label{sec:intro}
Suppose an investor intends to build one or more power plants to participate in an electricity market. The investor may already own units bidding in the same market. By adding generation capacity and depending on transmission congestion and load demand, electricity prices and generation schedules may be altered in a way so that its total financial gain from existing as well as new units is lowered. The goal of the investor is to find the optimal location and size of the new generation units to maximize its total profit. This task of \emph{strategic investment (SI)} is challenging for three reasons. First, the variables involved in SI, namely the generation schedules and prices, are not known beforehand but are computed as the solutions of an optimization problem, a linearized optimal power flow (OPF). Second, an investment can change the market outcome (prices and generation schedules), rendering SI a complex closed-loop problem. Third, increasing uncertainties introduced by renewable generation, loads, fuel prices, and bids from rival generators, call for stochastic methods thus further increasing the complexity of SI.

A promising method to handle the closed-loop complication of the SI task is posing it as a bilevel optimization~\cite{PSC17}. The inner level involves the OPF that clears the market and decides generation schedules and prices given generation capacities. The outer level aims to maximize the market profit for dispatching new and existing units minus the investment cost for the new units. Nonetheless, this bilevel formulation calls for complex complementarity methods~\cite{KCR12}; see~\cite{Investment} for a comprehensive survey. In the case of market OPFs with linear constraints, complementarity methods convert the bilevel problem into a single-level optimization upon replacing the inner problem with its Karush-Kuhn-Tucker (KKT) conditions. While complementarity methods promise globally optimal investment decisions, they entail computationally prohibitive mixed-integer programs. Such models may not scale gracefully in large power networks, and not be able to consider a sufficient number of scenarios. Alternatively, works like~\cite{LSC18},~\cite{MWJ15},~\cite{DKBP18} use scenario-based alternating direction method of multipliers or progressive hedging to decompose the related mixed-integer programs, at the expense of losing global optimality.

The complexity of complementarity models prompted us to waive the need for bilevel programming. The SI task could be dealt with by solving the OPF clearing the market for each possible combination of scenario and investment option. However, this process is also challenging due to the sheer number of OPFs that need to be solved, calling for efficient OPF solvers for a large number of market scenarios. Interestingly, the OPF problem under the linearized grid model, the so-called DC-OPF, can be viewed as an instance of multiparametric programming (MPP), where loads, generation capacities, and bids are considered as its parameters. As explained in Section~\ref{sec:mpp}, under certain conditions, the MPP toolbox can partition the parameter space into polytopes termed \emph{critical regions} for which the primal and dual solutions can be identified as affine functions of the problem parameters~\cite{BBM03},~\cite{TJB03}. The boundaries of these regions as well as the associated affine functions depend on which constraints are active at optimality, and hence, the DC-OPF needs to be solved only once per critical region. This latter property facilitates solving a large number of DC-OPFs with relatively small computational burden. 

MPP has been utilized before in power systems operations. The notion of congestion patterns in energy markets identified by~\cite{ZTL11} pertains exactly to the critical regions of MPP. The same regions also give rise to the active sets learned in~\cite{NMRB18}. References~\cite{JTT15} combined MPP with importance sampling over critical regions to compute the probability distribution of locational marginal prices (LMPs). The polytopic description of critical regions allows~\cite{GX17} to train a support vector machine classifier and estimate LMPs given loads. Reference~\cite{MBGT19} utilizes MPP and proposes a critical region exploration algorithm to solve a security-constrained economic dispatch. In the context of distribution grids, reference~\cite{TJKT20} leverages MPP to handle efficiently a large number of distribution OPF instances, and thus expedite probabilistic hosting capacity analysis. However, none of the previous works engages MPP to deal with the complex bilevel setup involved in SI. 

The contribution of this work is fourfold: \emph{c1)} Study SI analytically for a simple power network to demonstrate the challenges involved; \emph{c2)} Extend existing MPP claims to the OPF problem used to clear electricity markets; \emph{c3)} Develop an algorithm to compute efficiently the primal/dual outcomes of hundreds of OPF instances at a time. The algorithm can accelerate by an order of magnitude a brute-force grid search to cope with the SI task when the number of investment locations and scenarios are relatively small; and \emph{c4)} Devise a stochastic gradient descent (SGD) scheme to address directly the outer layer of the SI task, especially when multiple investment locations are considered. By uniquely exploiting MPP properties, this SGD scheme calculates gradients in a highly scalable mini-batch fashion.



\section{Strategic Investment in Electricity Markets}\label{sec:market}

\subsection{Modeling Electricity Markets}\label{subsec:model}
Suppose the energy market operates over a system with $N$ buses and $L$ transmission lines. In a wholesale electricity market, the independent system operator (ISO) calculates the generation schedule and electricity prices upon solving a linear or quadratic program to minimize the total generation cost subject to power balance and line flow constraints. From the viewpoint of a strategic investor, one can identify three types of generators~\cite{KCR12}: existing units owned by rival entities; existing units owned by the investor; and new units to be built by the investor. The power schedules corresponding to three unit types are denoted respectively by $(\bp_r,\bp_e,\bp_n)$. For notational brevity, suppose all unit types exist at all buses with possibly zero capacities. The market is cleared by the DC-OPF:
\begin{subequations}\label{eq:DCOPF}
	\begin{align}
		\underset{\bp_r,\bp_e,\bp_n}{\min}~&~f_r(\bp_r)+f_e(\bp_e)+f_n(\bp_n)\\
		\mathrm{s.to}~&~\mathbf 1^\top(\bp_r+\bp_e+\bp_n-\bdell)=0&&:\lambda_0\label{con:pbal}\\
		&~-\overline{\bef}\leq\bS(\bp_r+\bp_e+\bp_n-\bdell)\leq\overline{\bef}&&:\underline{\bmu},\overline{\bmu}\label{con:linelim}\\
		&~\bzero\leq\bp_r\leq\bbp_r&&:\underline{\bgamma}_r,\overline{\bgamma}_r\label{eq:pr-lim}\\
		&~\bzero\leq\bp_e\leq\bbp_e&&:\underline{\bgamma}_e,\overline{\bgamma}_e\label{eq:pe-lim}\\
		&~\bzero\leq\bp_n\leq\bbp_n&&:\underline{\bgamma}_n,\overline{\bgamma}_n\label{con:pn-lim}
	\end{align}
\end{subequations}
where $\bS$ is the power transfer distribution factor matrix~\cite{KGB16}. Function $f_r(\bp_r) := \frac{1}{2}\bp_r^\top\bH_r\bp_r+\bc_r^\top\bp_r$ models the generation cost for rival units. The diagonal matrix $\bH_r$ and vector $\bc_r$ contain positive values~\cite{WollenbergBook}. The generation costs for existing and new units $f_e$ and $f_n$ are defined similarly. Constraint~\eqref{con:pbal} ensures power balance with $\bdell$ being the vector of nodal load demands. Constraint \eqref{con:linelim} enforces given line flow limits $\overline{\bef}$. Constraints~\eqref{eq:pe-lim}--\eqref{eq:pr-lim} impose capacity limits $(\bbp_r,\bbp_e,\bbp_n)$ on generation schedules. Dual variables are shown in the right-hand side of the constraints in \eqref{eq:DCOPF}. 

To account for renewable generation, vector $\bbp_r$ is the \emph{available} capacity of rival units. It can be modeled as $\bbp_r=\balpha_r\odot\hbp_r$, where $\hbp_r$ is the vector of installed capacities and $\balpha_r$ the vector of capacity factors. For non-renewable generators, the corresponding entry of $\balpha_r$ is unity, whereas for renewable generators it changes with time to capture the available wind energy as a percentage of the maximum capacity. We similarly define vectors $(\balpha_e,\hbp_e)$ for existing units, and $(\balpha_n,\bx=\hbp_n)$ for new units. We will be using $\bx$ instead of $\hbp_n$ to emphasize that the capacity of new units is the ultimate optimization variable for the SI task at hand. In other words, the investor would eventually build new generation capacities $\bx$.

The ISO solves~\eqref{eq:DCOPF} every hour to find the optimal schedules $(\bp_r,\bp_e,\bp_n)$ and computes the locational marginal prices (LMPs) of electricity across buses as
\begin{equation}\label{eq:lmps}
\bpi = -\lambda_0\bone+\bS^\top(\underline{\bmu}-\overline{\bmu}).
\end{equation}
We have slightly abused notation and used the same symbols with \eqref{eq:DCOPF} to denote the optimizers of the problem. 
We next present SI adapting the formulation of~\cite{KCR12}.

\subsection{Problem Formulation}\label{subsec:problem}
Strategic investment in electricity markets can be viewed as a minimization problem where the objective is the amortized cost for investing in the new units minus the expected revenue obtained from the market through the new units and the existing own units. The investment cost is generally a known linear function $\bk^\top\bx$ of the generation capacity. The revenue is made up by the payment received from the ISO (generation schedule times LMP) minus the true generation cost
\begin{equation}\label{eq:fun}
f(\bx) := \bk^\top\bx-\mathbb{E}\left[\bpi^\top\left(\bp_e+\bp_n\right)-g_e(\bp_e)-g_n(\bp_n)\right]
\end{equation}
where the expectation $\mathbb{E}[\cdot]$ applies over the involved uncertainties. Note $f$ involves the actual cost of generation $g_e(\bp_e)+g_n(\bp_n)$ rather than the bid $f_e(\bp_e)+f_n(\bp_n)$ submitted to the market. This is because the market bid can be sometimes larger than the actual generation cost for some or all $(\bp_e,\bp_n)$~\cite{KCR12}. The expectation in~\eqref{eq:fun} is applied over all random quantities, such as the demand vector $\bdell$, the scaling factors $(\alpha_r,\balpha_e,\balpha_n)$ for renewable generation, and possible changes in bids.

The task of strategic investment can be now stated as
\begin{subequations}\label{eq:si}
	\begin{align}
	\underset{\bx\in\mcX}{\min}~&~f(\bx)\label{eq:si:a}\\
	\mathrm{s.to}~&~\left\lbrace\bpi,\bp_e,\bp_n\right\rbrace~\textrm{being solutions of}~\eqref{eq:DCOPF}\label{eq:si:b}.
	\end{align}
\end{subequations}
In addition to constraint \eqref{eq:si:b}, the investment variable $\bx$ should also belong to the set of investment options $\mcX:=\{\bx:\underline{\bx}\leq \bx\leq \overline{\bx},~\bDelta\bx\leq \bdelta\}$. Constraint $\bDelta\bx\leq \bdelta$ could model an upper bound on the total MW capacity installed or the total number of wind turbines purchased. In this case, matrix $\bDelta$ degenerates to the all-one vector $\bDelta=\bone^\top$ and vector $\bdelta$ to a scalar $x_{\text{total}}$, where $x_{\text{total}}$ is the total capacity to be installed. Investments $\bx$ may also be restricted to take discrete values. As in~\cite{KCR12}, we further postulate two assumptions on the problem setup. 

\begin{assumption}\label{assum:Sknown}
The transmission network topology captured by $(\bS,\overline{\bef})$ is known and remains constant.
\end{assumption}

\begin{assumption}\label{assum:feas}
The problem parameters $(\bdell,\bbp_r,\bbp_e,\overline{\bef})$ are such that the DC-OPF of \eqref{eq:DCOPF} is feasible for all $\bx\in\mcX$. 
\end{assumption}

According to Assumption~\ref{assum:feas}, the power system can be dispatched without the new units. Even under these assumptions, problem~\eqref{eq:si} is challenging due to three reasons: \emph{i)} Constraint~\eqref{eq:si:b} is expressed as an optimization problem itself; \emph{ii)} The products between primal and dual variables inside the expectation in \eqref{eq:fun} are non-convex functions; and \emph{iii)} Evaluating the expectation in $f(\bx)$ may be prohibitive. The investment task of~\eqref{eq:si} will be termed the \emph{outer problem} and the DC-OPF of~\eqref{eq:DCOPF} given $\bx$ as the \emph{inner problem}. 


\begin{figure}[t]
	\centering
	\includegraphics[width=0.3\textwidth]{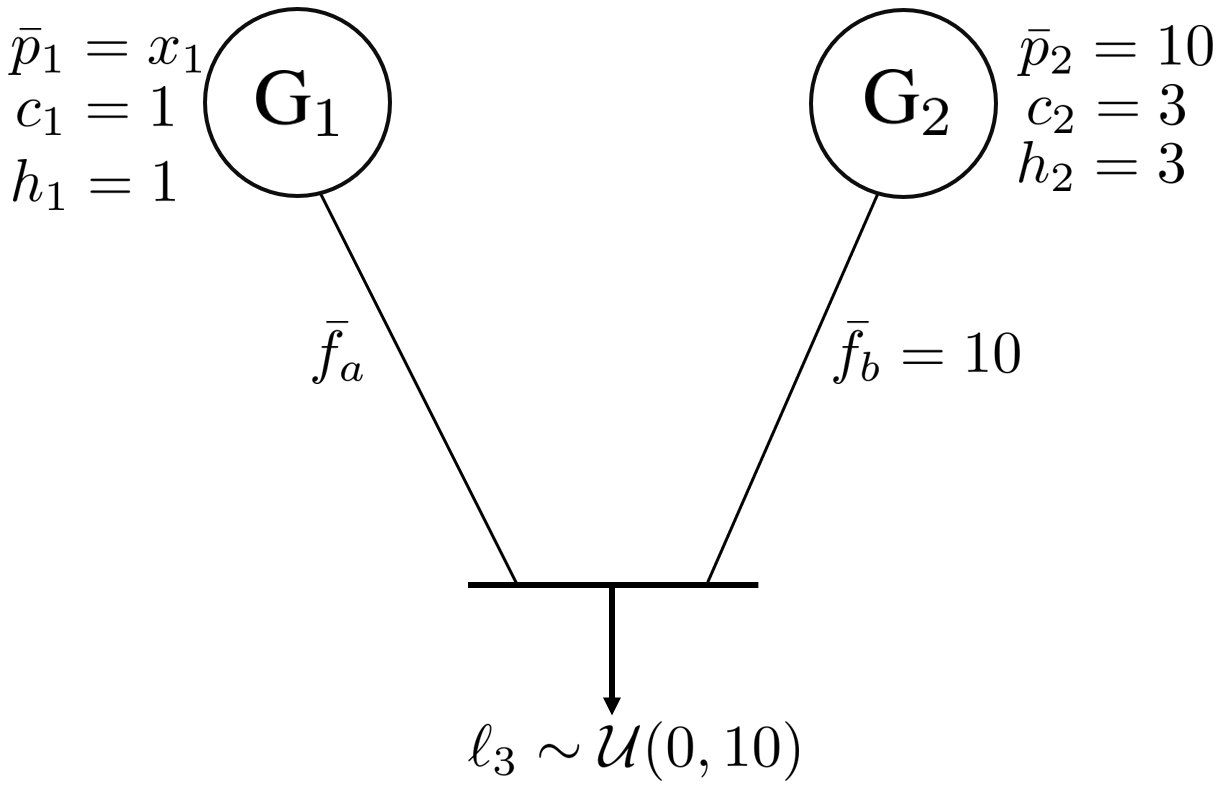}
	\caption{A 3-bus power system showcasing the challenges in solving~\eqref{eq:si}.}
	\label{fig:3bus}
\end{figure}

\begin{figure}[t]
	\centering
	\includegraphics[width=0.4\textwidth]{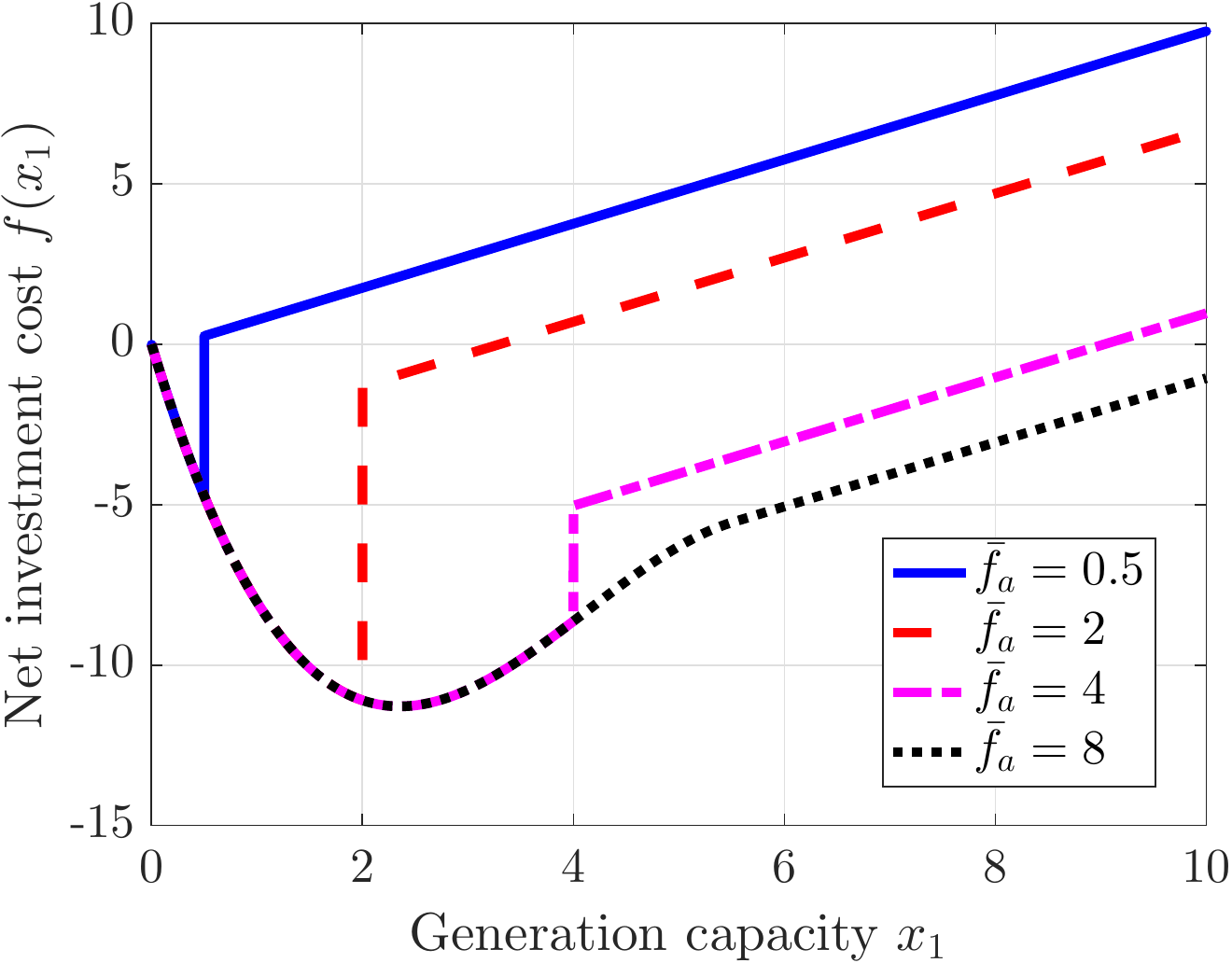}
	\vspace*{-.5em}
	\caption{The net investment cost for the system on Figure~\ref{fig:3bus} over the capacity range of $x_1\in\left[0,10\right]$ and for different capacities $\bar{f}_a$ of line $a=(1,3)$.}
 	\label{fig:3bus-cost}
 	\vspace*{-2em}
\end{figure}

SI can be challenging due to non-convexity and discontinuities. To elucidate on that, consider the 3-bus power system of Fig.~\ref{fig:3bus}. An investor considers building a generator on bus 1. A rival generator is located at bus 2 having capacity $\bar{p}_2=10$~pu. Bus 3 hosts a load $\ell_3$ whose value is modeled as a random variable uniformly distributed within $(0,10)$. The limit for line $b=(2,3)$ is $\bar{f}_b=10$, while the limit for line $a=(1,3)$ is left as a variable $\bar{f}_a$ to study its effect on SI. 
The investor does not own an existing generator, so that $f_e(\bp_e)=g_e(\bp_e)=0$. Let us assume quadratic bidding functions $f_n(\bp_n)=p_1^2+p_1$ and $f_r(\bp_r)=p_2^2+3p_2$; investment cost $k_1=1$; and $g_n(\bp_n)=f_n(\bp_n)$. Given $\ell_3$ is uniformly distributed, the expectation in~\eqref{eq:si:a} can be evaluated. Sparing derivations due to space limitations, Figure~\ref{fig:3bus-cost} plots the investment cost of~\eqref{eq:fun} for different values of $\bar{f}_a$. 


\section{Preliminaries}\label{sec:prelims}
This sections reviews mathematical programming with equilibrium constraints (MPEC) and multiparametric programming (MPP), on which our methodology builds upon.

\subsection{Mathematical Programming with Equilibrium Constraints}\label{sec:mpec}
MPEC is used in economics, where decisions taken by an investor affect the outcome of a market. MPEC results at bilevel optimization programs, such as the one in~\eqref{eq:si}. Reference~\cite{KCR12} posed \eqref{eq:si} as an MPEC. For later reference, we derive a formulation similar to the one in~\cite{KCR12}, but for markets with quadratic bidding costs. The model is built in three steps.

First, the bilinear term $\bpi^\top(\bp_e+\bp_n)$ in \eqref{eq:fun} is replaced by a quadratic function of the variables of the inner problem
\begin{align}\label{eq:prod}
\bpi^\top(\bp_e+\bp_n)&=-\bp_r^\top\bH_r\bp_r-\bc_r^\top\bp_r - (\overline{\bmu} +\underline{\bmu})^\top\overline{\bef} -\overline{\bgamma}_r^\top\bbp_r\nonumber\\
&\quad- (\overline{\bmu} -\underline{\bmu})^\top\bS\bdell -\lambda_{0}\bone^\top\bdell.
\end{align}
This follows from the strong duality of the inner problem and after some algebraic manipulations, which are omitted due to space limitations. For linear bidding functions, a similar result can be obtained by setting $\bH_r=\bzero$~\cite{KCR12}. 

As a second step, the expectation in \eqref{eq:fun} is surrogated by a sample average over $T$ scenarios indexed by $t$ as~\cite{KCR12}
\begin{equation*}
\hat{f}(\bx)=\bk^\top\bx-\frac{1}{T}\sum_{t=1}^T\bpi_t^\top(\bp_{e,t}+\bp_{n,t})-g_e(\bp_{e,t})-g_n(\bp_{n,t})
\end{equation*}
so the SI problem can be approximated as
\begin{align}\label{eq:sih}
\underset{\bx\in\mcX}{\min}~&\hat{f}(\bx)\\
\mathrm{s.to}~&\left\lbrace\bpi,\bp_{e,t},\bp_{n,t}\right\rbrace~\textrm{being solutions of}~\eqref{eq:DCOPF}~\textrm{for}~t=1:T.\nonumber
\end{align}
	
The third step replaces constraint \eqref{eq:si:b} by the KKT conditions for the inner problem. Primal/dual feasibility and Lagrangian optimality yield a set of linear equality and inequality constraints on the primal/dual variables of the inner problem. Complementary slackness conditions entail non-convex products between primal and dual variables, which can be encoded via the big-M method~\cite{KCR12}. For example, the complementary slackness condition $x\cdot\lambda=0$ for a primal constraint $x\geq 0$ and the related Lagrange multiplier $\lambda \geq 0$ can be written as
\begin{equation}\label{eq:bigm}
0\leq x\leq \phi M\quad \quad 0\leq \lambda \leq (1-\phi)M
\end{equation}
where $M$ is a large constant and $\phi$ is an auxiliary binary variable. A set of constraints similar to \eqref{eq:bigm} has to be enforced for each constraint of the inner problem and every market scenario $t$. Following the aforesaid three steps, the bilevel problem in~\eqref{eq:si} can be reformulated as a mixed-integer linear or quadratic program (MILP/MIQP), depending on whether functions $\{f_r,f_e,f_n,g_e,g_n\}$ are linear or quadratic.

The MPEC method of \cite{KCR12} finds the global minimum of \eqref{eq:si} after approximating the expectation with scenarios. Nonetheless, the resultant mixed-integer model may not scale favorably for large $T$ or large networks. Moreover, finding proper values for $M$ is challenging, since it is hard to upper bound dual variables \emph{a-priori}. To avoid computationally taxing mixed-integer models, we develop solvers of \eqref{eq:si} leveraging the powerful tool of MPP, which is outlined next.

\subsection{Multiparametric Programming (MPP)}\label{sec:mpp}
MPP is a tool for characterizing the solutions of optimization problems dependent on a parameter vector~\cite{BBM03}. The main idea of using MPP for SI is to handle the inner problem in \eqref{eq:si} as an MPP with the SI vector $\bx$ as its parameter. To this end, consider a minimization dependent on parameter $\btheta$ as
\begin{subequations}\label{eq:qp}
	\begin{align}
	\underset{\bp}{\min}~&~\frac{1}{2}\bp^\top\bH\bp+\left(\bC\btheta+\bd\right)^\top\bp\label{eq:qp:cost}\\
	\mathrm{s.to}~&~\bA\bp\leq\bE\btheta+\bb&&:\blambda\label{eq:qp:ineq}\\
	&~\bB\bp=\bF\btheta+\by.&&:\bmu.\label{eq:qp:eq}
	\end{align}
\end{subequations}
If $\bH=\bzero$, problem \eqref{eq:qp} is a multiparametric linear program (MPLP). If $\bH\succeq\bzero$, problem \eqref{eq:qp} is a multiparametric convex quadratic program (MPQP). 

Let $\Theta$ be the set of all $\btheta$'s for which \eqref{eq:qp} is feasible. According to the MPP theory~\cite{BBM03}, set $\Theta$ can be partitioned into distinct regions, termed \emph{critical regions}, with three interesting properties: \emph{p1)} Each region is described as a polytope in $\Theta$; \emph{p2)} Within each region, the same subset of inequality constraints become \emph{active}, i.e., are satisfied with equality; and \emph{p3)} Within each region, the primal/dual solutions of \eqref{eq:qp} can be expressed as affine functions of $\btheta$. These affine functions have been derived in~\cite{TJKT20}. They are reviewed next for completeness and to introduce the needed quantities.

Assume \eqref{eq:qp} is solved for $\btheta_o\in\Theta$ and let $\left(\bp_o;\blambda_o,\bmu_o\right)$ be the obtained optimal primal/dual solutions. Let also $\tbA$ be the submatrix obtained from $\bA$ upon selecting the rows corresponding to the active constraints in~\eqref{eq:qp:ineq}. The remaining rows of $\bA$ related to \emph{inactive constraints} (constraints satisfied with strict inequality) constitute matrix $\bbA_o$. Similar partitions yield $(\tbE,\tbb,\tblambda)$ and $(\bbE,\bbb,\bblambda)$. It is further assumed that matrix
\begin{equation}\label{eq:Kmat}
\bK:=[\tbA^\top~~\bB^\top]^\top
\end{equation}
is full row-rank. This condition is known as \emph{linear independence constraint qualification (LICQ)}. Although LICQ cannot be guaranteed before solving \eqref{eq:qp} for a $\btheta$, it occurs in the majority of our tests in Section~\ref{sec:tests}. We next consider separately the cases of $\bH\succ\bzero$ and $\bH=\bzero$ in \eqref{eq:qp}. 

For $\bH\succ\bzero$ (strictly convex MPQP) and under LICQ, the primal/dual solutions of \eqref{eq:qp} can be obtained as~\cite{TJKT20}
\begin{align}\label{eq:MPP_sol}
\begin{bmatrix}
\bp_o\\
\tblambda_o\\
\bmu_o
\end{bmatrix} =\bM\btheta_o+\br=\begin{bmatrix}
\bM_1\\
\bM_2\\
\bM_3
\end{bmatrix}\btheta_o+\begin{bmatrix}
\br_1\\
\br_2\\
\br_3
\end{bmatrix}
\end{align}
where
\begin{subequations}\label{eq:MPP-q}
\begin{align}
&\bM=\begin{bmatrix}
\bM_1\\\bM_2\\\bM_3\end{bmatrix}:=\begin{bmatrix}
\bH&\tbA^\top&\bB^\top\\\tbA&\bzero&\bzero\\\bB&\bzero&\bzero\end{bmatrix}^{-1}\begin{bmatrix}-\bC\\\tbE\\\bF\end{bmatrix}\label{eq:M}\\
&\br=\begin{bmatrix}
\br_1\\\br_2\\\br_3\end{bmatrix}:=\begin{bmatrix}
\bH&\tbA^\top&\bB^\top\\\tbA&\bzero&\bzero\\\bB&\bzero&\bzero\end{bmatrix}^{-1}\begin{bmatrix}\bd\\\tbb\\\by\end{bmatrix}\label{eq:r}.
\end{align}
\end{subequations}
The matrix inverse in~\eqref{eq:MPP-q} exists, since its determinant equals $\det(\bH)\det(-\bK\bH^{-1}\bK^\top)<0$ from Schur's complement. 

For $\bH=\bzero$ (for which \eqref{eq:qp} is an MPLP), suppose further that $\bK$ is square. This holds if in addition to LICQ, the number of active constraints equals the number of optimization variables. Then, the primal/dual solutions of \eqref{eq:qp} take again the closed-form expression of \eqref{eq:MPP_sol}, but with~\cite{TJKT20}
\begin{subequations}\label{eq:MPP-l}
\begin{align}
&\bM_1:= \bK^{-1}\begin{bmatrix}\tbE\\\bF\end{bmatrix}\quad\textrm{and}\quad\br_1:= \bK^{-1}\begin{bmatrix}\tbb\\\by\end{bmatrix}\label{eq:M-lin1}\\
&\begin{bmatrix}\bM_2\\\bM_3\end{bmatrix}:=\bK^{-\top}\bC\quad\textrm{and}\quad
\begin{bmatrix}\br_2\\\br_3\end{bmatrix}:=\bK^{-\top}\bd\label{eq:r-lin2}.
\end{align}
\end{subequations}

One of the interesting claims of MPLP/MPQPs is that for any other $\btheta\in\Theta$ yielding rise to the same set of active constraints, the primal/dual solutions are expressed through \eqref{eq:MPP_sol}; see e.g.,~\cite{BBM03}, \cite{TJB03}. Contrarily, given a set of constraints, the subset of $\btheta$'s activating those constraints can be identified as a polytope $\mcC\subseteq\Theta$ described as (see \cite{TJB03} for details)
\begin{equation}\label{eq:MPP-region}
\mcC :=\left\{\btheta\in\Theta|\left(\bbA\bM_1-\bbE\right)\btheta\leq\bbb-\bbA\br_1,\bM_2\btheta\geq\br_2\right\}.
\end{equation}
The quantities $(\bM_1,\bM_2,\br_1,\br_2)$ are provided by \eqref{eq:MPP-q} or \eqref{eq:MPP-l} for MPQP and MPLP, accordingly. The set $\mcC$ is termed a \emph{critical region} of $\Theta$. We next leverage these MPP properties to cope with~\eqref{eq:si} in two different ways.

\section{Strategic Investment via MPP}\label{sec:mpp4si}
To derive efficient SI solvers, the key idea is to cast the inner problem as an MPP and exploit the rich properties for its solutions. If the bidding functions in~\eqref{eq:DCOPF} are quadratic or (piecewise) affine, then \eqref{eq:DCOPF} is an instance of~\eqref{eq:qp}. The optimization variable $\bp$ stacks the variables  $(\bp_r,\bp_e,\bp_n)$. The parametric inequalities of~\eqref{eq:qp:ineq} capture the line flow constraints of~\eqref{con:linelim} and the generation limits of~\eqref{eq:pr-lim}--\eqref{eq:pe-lim}. The parametric equalities~\eqref{eq:qp:eq} relate to the power balance constraint of~\eqref{con:pbal}.  

The parameter $\btheta$ appearing in \eqref{eq:qp} consists of three parts. The first part relates to varying generation cost coefficients (bids). Under the assumption that the quadratic component $\frac{1}{2}\bp^\top\bH\bp$ remains invariant across scenarios, these changing costs are modeled by $\bC\btheta+\bd$ in~\eqref{eq:qp:cost} with $\bC=[\bI ~\bzero~\bzero]$ and $\bd=\bzero$. The second part of $\btheta$ captures the uncertain demand vector $\bdell$. The third part captures varying capacities of generation units that are due to scheduled outages or due to variable renewable resources. Moreover, the capacity for new units will be changing while solving SI as solvers will be evaluating \eqref{eq:DCOPF} for different values of $\bx$ seeking the optimal investment. In summary, the parameter vector $\btheta$ can be expressed as
\begin{equation}\label{eq:theta}
\btheta :=[\bc^\top~\bdell^\top~\bbp^\top]^\top
\end{equation}
where $\bc :=[\bc_r^\top~\bc_e^\top~\bc_r^\top]^\top$; $\bbp :=[\bbp_r^\top~\bbp_e^\top~(\balpha\odot\bx)^\top]^\top$; and $\balpha :=[\balpha_r^\top~\balpha_e^\top~\balpha_r^\top]^\top$. Collect all uncertain components of $\btheta$ in $\bomega:=\{\bc_r,\bc_e,\bc_r,\bdell,\bbp_r,\bbp_e,\balpha\}$. Evidently from~\eqref{eq:theta}, the parameter vector of \eqref{eq:qp} can be expressed as a mapping $\btheta=\mcP(\bomega,\bx)$ of the uncertain variables $\bomega$ and the optimization variable of the outer problem $\bx$. Heed that the mapping $\mcP(\bomega,\bx)$ is not linear in $(\bomega,\bx)$ due to the products $\balpha\odot\bx$ in $\bbp$. Nonetheless, the objective and constraint functions of the parametric QP in \eqref{eq:qp} depend \emph{linearly} on $\btheta=\mcP(\bomega,\bx)$. The matrices $(\bA,\bE,\bB,\bF)$ and vectors $(\bb,\by)$ in \eqref{eq:qp} are straightforward to compute and are not presented here. Having posed~\eqref{eq:DCOPF} as an instance of~\eqref{eq:qp}, we next present two methods that leverage the affine mappings of~\eqref{eq:MPP_sol}--\eqref{eq:MPP-l} and the partitioning of~\eqref{eq:MPP-region} to solve \eqref{eq:sih}.


\subsection{An MPP-aided Grid Search (MPP-GS) Scheme}\label{subsec:MPP-GS}
This section exploits the MPP toolbox of Section~\ref{sec:mpp} to solve \eqref{eq:DCOPF} for a large number of $(\bomega,\bx)$ instances. We can thus evaluate $\hat{f}(\bx)$ over a grid of $\bx$ values efficiently. This grid search approach is preferred when an investor is presented with a single or few possible investment locations. To design our search grid, note that the investment $x_m$ at bus $m$ can be bounded as
\begin{equation}\label{eq:xmax}
\overline{x}_m = \sum_{k\sim m}\overline{f}_{m,k}
+\max_{t}\{\ell_{m,t}\}
\end{equation}
by the maximum load at bus $m$ plus the sum of capacities for all transmission lines incident to bus $m$. Symbol $\overline{f}_{m,k}$ denotes the capacity of the line connecting buses $m$ and $k$, if such line exists. The quantity $\overline{x}_m$ is the maximum power that can be produced at bus $m$ without violating any physical limits. The  discretization step over $\left[0,\bar{x}_m\right]$ can be chosen based on the type of the power plant. For example, a typical wind turbine is about 2-3~MW, so that multiples of this value are reasonable options for the grid step. When investing at $M$ locations with $K_m$ search values per location $m$, we get a search grid $\hat{\mcX}\subseteq \mcX$ of $K=\prod_{m=1}^M K_m$ points. Since $K$ grows exponentially with $M$, this approach makes sense only for $M=1-3$ locations. 

Given the search grid $\hat{\mcX}$ and the uncertain parameter set $\Omega:=\{\bomega_t\}_{t=1}^T$, one can readily form the parameter set $\hat{\Theta}$ using the mapping $\mcP:\hat{\mcX}\times \Omega\rightarrow \hat{\Theta}$ and $|\hat{\Theta}|=KT$. Here $\hat{\Theta}$ is a finite subset of $\Theta$, over which \eqref{eq:qp} has to be solved. This slightly abuses notation since in Section~\ref{sec:mpp} symbol $\Theta$ denoted the convex set of $\btheta$'s rendering \eqref{eq:qp} feasible. A solution to~\eqref{eq:sih} can be found by solving \eqref{eq:DCOPF} in its parameterized form of \eqref{eq:qp} for all $KT$ members of $\hat{\Theta}$, and then evaluating $\hat{f}(\bx)$ over $\hat{\mcX}$. For $\hat{f}(\bx)$ to be a reasonable estimate of $f(\bx)$ though, a large number $T$ of scenarios $\bomega_t$ needs to be considered, yielding a computationally formidable task even for small $K$. 

Thanks to MPP however, problem~\eqref{eq:qp} needs to be solved for just as many times as the critical regions appearing in $\hat{\Theta}$. To see this, suppose that for a critical region $\mcC_o\subseteq\Theta$, we have already computed its polytopic description in \eqref{eq:MPP-region} and the pair $(\bM,\br)$ parameterizing its primal/dual solutions. Then, for any other $\btheta_s\in \hat{\Theta}$ belonging to $\mcC_o$, we can directly compute its primal/dual solutions from \eqref{eq:MPP_sol} without having to solve \eqref{eq:qp}. This procedure, termed \emph{MPP-based Grid Search (MPP-GS)}, is formalized as Algorithm~\ref{alg:SGS} and its steps are explained next. 

\begin{algorithm}[t]
	\caption{MPP-aided Grid Search (MPP-GS)}\label{alg:SGS}
	\begin{algorithmic}[1]
		\renewcommand{\algorithmicrequire}{\textbf{Input:}}
		\renewcommand{\algorithmicensure}{\textbf{Output:}} 
		\REQUIRE Set of OPF scenarios $\hat{\Theta}=\left\{\btheta_s\right\}_{s=1}^{KT}$
		\ENSURE OPF solutions $\{\bpi_s,\bp_{e,s},\bp_{n,s}\}_{s=1}^{KT}$ to \eqref{eq:DCOPF} via \eqref{eq:qp} for all $\btheta_s\in\hat{\Theta}$
		\WHILE {$\hat{\Theta}\neq \emptyset$}
		\STATE Randomly select $\btheta_o\in\hat{\Theta}$ and $\hat{\Theta}\gets\hat{\Theta}\setminus\btheta_o$
		\STATE Solve~\eqref{eq:qp} for $\btheta_o$ to find its primal/dual solutions  and active constraints
		\STATE Record $(\bp_{e,o},\bp_{n,o},\bpi_o)$
		\IF {matrix $\bK$ of \eqref{eq:Kmat} is full row-rank,}
		\STATE Compute region's parameters $(\bM,\br)$ from \eqref{eq:MPP_sol}
		\STATE Compute region's polytope $\mcC$ from \eqref{eq:MPP-region}
		\FOR {all $\btheta_s\in\hat{\Theta}$}
		\IF {$\btheta_s\in\mcC$ [satisfying \eqref{eq:MPP-region}],}
		\STATE Compute OPF solution as $\bp_s=\bM\btheta_s+\br$
		\STATE Record $(\bp_{e,s},\bp_{n,s},\bpi_s)$
		\STATE $\hat{\Theta}\gets\hat{\Theta}\setminus\btheta_s$ 
		\ENDIF
		\ENDFOR
		\ENDIF
		\ENDWHILE
		\STATE Evaluate $\hat{f}(\bx)$ over $\hat{\mcX}$ and find the minimizing $\bx$
	\end{algorithmic}
\end{algorithm}

MPP-GS selects a $\btheta_o$ from $\hat{\Theta}$ at step 2. At step 3, it solves \eqref{eq:qp} for $\btheta_o$. If the related $\bK$ is of full row rank, the algorithm constructs a description for the visited critical region (steps 6-7). It further scans the remaining dataset $\hat{\Theta}$ to find other $\btheta_s$'s belonging to this region (step 8); computes their solution in closed form (steps 10-11); and removes these $\btheta_s$'s from $\hat{\Theta}$ (step 12). The process continues until $\hat{\Theta}$ becomes empty.

MPP-GS explores a critical region only when $\bK$ is of full row rank (step 5). 
Albeit such cases could be handled~\cite{BBM03,TJB03}, they involve methods of high complexity. Instead, when we come across such an instance of \eqref{eq:qp}, we only record its primal/dual solutions. During the tests of Section~\ref{sec:tests}, these instances appear infrequently. Vectors $\btheta_s$ are visited in an arbitrary rather than sequential fashion, by randomly sampling from $\hat{\Theta}$ (step 2). In this way, we increase the chances of exploring more popular critical regions early on. It is hence more likely to handle a larger number of $\btheta_s$'s earlier, so that $\hat{\Theta}$ shrinks faster and step 9 is run on progressively much fewer $\btheta_s$'s. 
To cope with \eqref{eq:sih} for larger $K$ or $T$, we next pursue an MPP-aided stochastic gradient descent approach.

\subsection{MPP-aided Stochastic Gradient Descent (MPP-SGD)}\label{subsec:MPP-SGD}
The objective $\hat{f}(\bx)$ of \eqref{eq:sih} involves a summation over a large number $T$ of scenarios $\bomega_t$. Rather than finding the costly gradient of $\hat{f}(\bx)$, we adopt stochastic approximation and update $\bx$ by taking each time a descent step over the gradient for only one of the summands of $\hat{f}(\bx)$. Define the summand of $\hat{f}(\bx)$ related to scenario $\bomega_t$ as
\begin{equation}\label{eq:ft}
f_t(\bx):=\bk^\top\bx-\bpi_t^\top(\bp_{e,t}+\bp_{n,t})+g_e(\bp_{e,t})+g_n(\bp_{n,t}).
\end{equation}
Recall that the dispatches $(\bp_{e,t},\bp_{n,t})$ and prices $\bpi_t$ are all functions of $\bx$, since they are outcomes of \eqref{eq:DCOPF} given $\bx$. 

Apparently $\nabla_\bx (\bk^\top\bx)=\bk$. To study the differentiability of the remaining terms of $f_t$, assume for now that $\btheta=\mcP(\bomega_t,\bx)$ is strictly inside a critical region $\mcC_o\subseteq\Theta$.
According to~\eqref{eq:MPP_sol}, the optimal dispatch vectors $\bp_{e,t}$ and $\bp_{n,t}$ are affine in $\btheta$ and hence, affine in $\bx$ for a particular $\bomega_t$.  to \eqref{eq:lmps}, optimal prices $\bpi_{t}$ are affine in $(\blambda,\bmu)$. Since $(\blambda,\bmu)$ are affine functions of $\btheta$ from \eqref{eq:MPP_sol}, the prices $\bpi_{t}$ are affine in $\btheta$ as well. Consequently, the revenue term $\bpi_t^\top\left(\bp_{e,t}+\bp_{n,t}\right)$ is quadratic in $\bx$ and its gradient takes the form
\begin{equation}\label{eq:prod-grad}
 \nabla_\bx\left[\bpi_t^\top\left(\bp_{e,t}+\bp_{n,t}\right)\right] = \bQ_o\btheta+\bq_o.
\end{equation}
The parameters $(\bQ_o,\bq_o)$ can be computed using~\eqref{eq:MPP_sol}. Heed these parameters remain \emph{constant} within each critical region of $\Theta$, that is for all pairs $(\bomega_t,\bx)$ for which $\btheta=\mcP(\bomega_t,\bx)\in\mcC_o$. As in Section~\ref{subsec:MPP-GS}, the uncertain parameters $\bomega_t$ are drawn from a finite set of scenarios. On the contrary, the investment variable $\bx$ is drawn now from a continuous set.

\begin{figure}[t]
    \centering
	\includegraphics[width=0.36\textwidth]{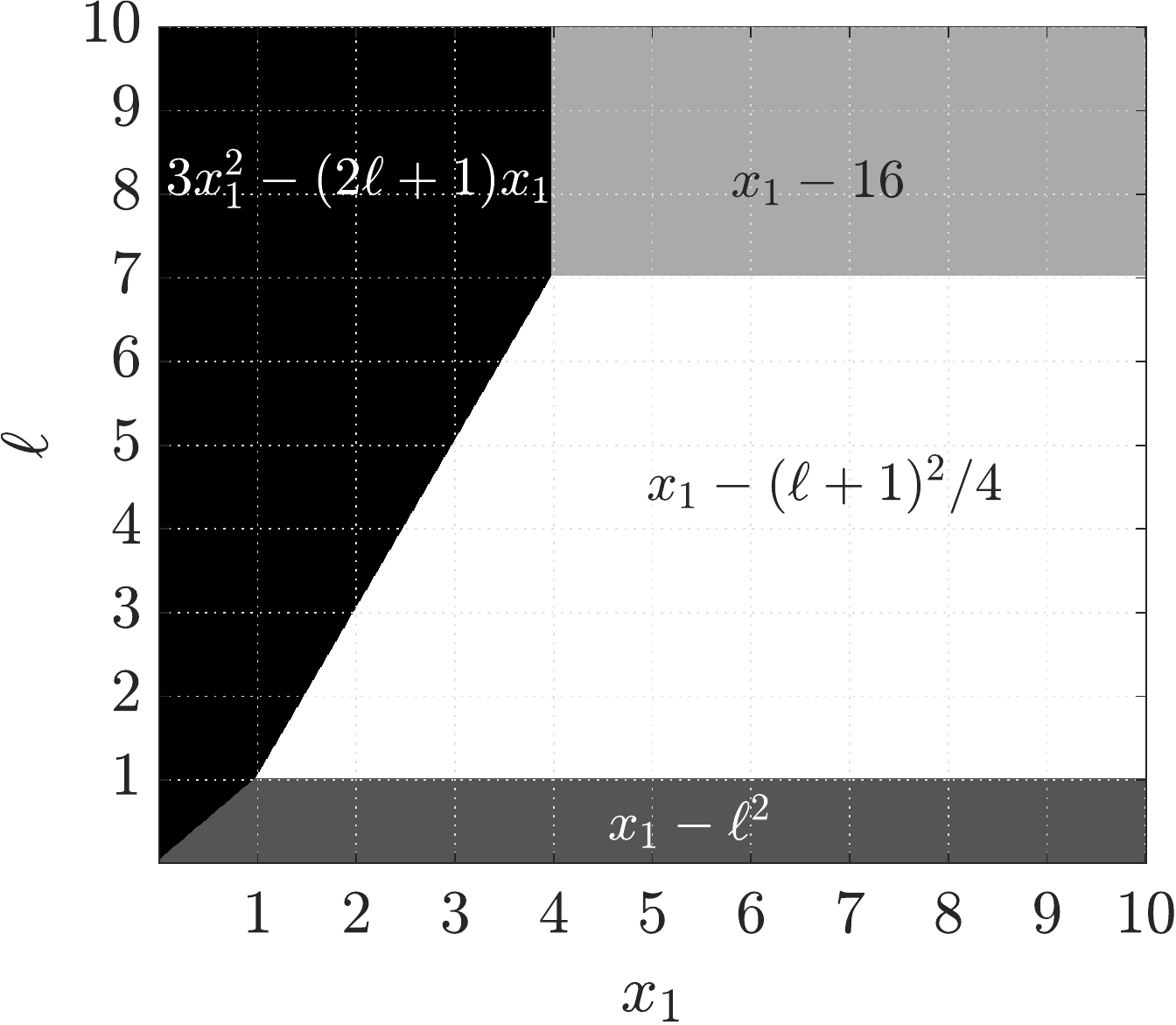}
	\vspace*{-1.0em}
	\caption{The critical regions of the inner problem in Figure~\ref{fig:3bus-cost} for $\bar{f}_a=4$. The dark gray region corresponds to $p_2$ at its lower limit ($p_2=0$); the black region to $p_1$ at its upper limit $p_1=x_1$; the light gray region to line $(1,3)$ being congested; and the white region to no active constraints.}
	\label{fig:fhat-grad}
 	\vspace*{-1.5em}
\end{figure}

Regarding the term $g_e(\bp_{e,t})+g_n(\bp_{n,t})$, its gradient with respect to $\bx$ can be computed using the chain rule, since functions $(g_e,g_n)$ are known (quadratic or affine) and $(\bp_{e,t},\bp_{n,t})$ are affine functions of $\btheta$ and consequently $\bx$. Consider now the case where $\btheta=\mcP(\bomega_t,\bx)$ is on a boundary between critical regions. Then, functions $\bpi_t^\top\left(\bp_{e,t}+\bp_{n,t}\right)$, $g_e(\bp_{e,t})$, and $g_n(\bp_{n,t})$ may not be differentiable or even continuous with respect to $\bx$. Take for example the 3-bus example of Figure~\ref{fig:3bus-cost}: Figure~\ref{fig:fhat-grad} shows its critical regions over $\btheta$. It also displays the functional form of $f_t(x_1)$ per region. Evidently, function $f_t(x_1)$ is differentiable within each region, but not on their boundaries. Nonetheless, these boundaries are zero-probability events over $\Theta$. Being a stochastic algorithm, the probability of coming across such $\btheta$'s during the SGD iterations is zero. 

Instead of updating $\bx$ for one $\bomega_t$ at a time, we exploit the MPP toolbox and derive a mini-batch rendition to get improved algorithmic convergence at a minimal increase in computational complexity. The novel idea is to exploit MPP regions and efficiently compute gradients with respect to $\bx$ not only for a single, but multiple $\bomega_t$'s at a time. To elaborate, notice that for a particular $\bx_o$, all $\btheta_t=\mcP(\bomega_t,\bx_o)$ that belong to the same critical region share the same gradient coefficients $(\bQ_o,\bq_o)$ in~\eqref{eq:prod-grad}. Hence, all these gradients can be readily computed once this critical region, its parameters $(\bM,\br)$, and its polytopic description of~\eqref{eq:MPP-region} have been identified.

Our MPP-aided stochastic gradient descent algorithm is tabulated as Algorithm~\ref{alg:MPP-SGD}. Step 3 constructs a parameter set $\Theta^k$ based on the current estimate of the investment vector $\bx^{k}$ and all scenarios $\bomega_t$'s. In steps 4-8, a random $\btheta_o$ is drawn from $\Theta^k$ and we identify the region it belongs to. Steps 9-14 compute the gradient with respect to $\bx$ for all $\btheta\in\Theta^{k}$ and sum them up in $\bg^{k}$. Step 12 counts the members of the said region, so that the average gradient can be computed in step 16. The updates of step 17 are guaranteed to converge to a stationary point~\cite{nemirovski2009robust}. The random draw of step 4 ensures an unbiased exploration of regions in $\Theta$, hence the average gradient per region is an unbiased estimate of the gradient of~$\hat{f}(\bx)$ in \eqref{eq:sih}.

\begin{algorithm}[t]
	\caption{MPP-Stochastic Gradient Descent (MPP-SGD)}\label{alg:MPP-SGD}
	\begin{algorithmic}[1]
		\renewcommand{\algorithmicrequire}{\textbf{Input:}}
		\renewcommand{\algorithmicensure}{\textbf{Output:}} 
		\REQUIRE $\Omega$, initialization $\bx_o$, tolerance $\tau$, and step size $\eta$
		\ENSURE Optimal investment $\bx^*$
		\STATE Set $\bx^0 = \bx_o$, $\epsilon>\tau$, $k=0$
		\WHILE {$\epsilon\geq\tau$}
		\STATE Define $\Theta^{k}\gets\{\btheta_1^{k},\ldots,\btheta_T^k\}$ where $\btheta_t^{k} = \mcP(\bomega_t,\bx^{k})$
		\STATE Randomly select $\btheta_o$ from $\Theta^k$
		\STATE Solve~\eqref{eq:qp} for $\btheta_o$ to find its primal/dual solutions  and active constraints
		\STATE Set $\bg^{k}\gets\bzero$ and $c^{k} \gets 0$
		\IF {matrix $\bK$ is full row-rank,}
		\STATE Compute region's parameters $(\bM,\br)$ from \eqref{eq:MPP_sol} and gradient coefficients $(\bQ_o,\bq_o)$
		\STATE Compute region's polytope $\mcC$ from \eqref{eq:MPP-region}
		\FOR {all $\btheta_t^k\in\Theta^{k}$}
		\IF {$\btheta_t^k\in\mcC$,}
		\STATE compute the gradient $\bg_t$ and set $\bg^{k}\gets\bg^{k}+\bg_t$
		\STATE set $c^{k}\gets c^{k}+1$
		\ENDIF
		\ENDFOR
		\ENDIF
		\STATE Set $\bx^{k+1} \gets \left[\bx^k-\frac{\eta}{c^{k}\sqrt{k}}\bg^k\right]_{\mcX}$
		\STATE Compute the moving average $\bbx^{k}\gets\frac{\sum_{i=\lceil k/2\rceil}^k\left(\bx^{(i)}/\sqrt{i}\right)}{\sum_{i=\lceil k/2\rceil}^k\sqrt{i}}$ 
		\STATE Set $\epsilon\gets\frac{\|\bbx^{k}-\bbx^{(k-1)}\|}{\|\bbx^{k}\|}$ and $k\gets k+1$
		\ENDWHILE
		\STATE Set $\bx^*=\bbx^k$
	\end{algorithmic}
\end{algorithm}

\section{Numerical Tests}\label{sec:tests}
Algorithms~\ref{alg:SGS} and \ref{alg:MPP-SGD} were contrasted against the MPEC method of \cite{KCR12} on three systems: the 3-bus system of Figure~\ref{fig:3bus-cost}; the IEEE 30-bus system; and the IEEE 118-bus system with line limits estimated from surge impedances per~\cite{Glover}. The investing bus were chosen as $\{1\}$, $\{3,20\}$, and $\{29,95\}$ for the three networks, respectively. For the 30- and 118-bus systems, we chose the investor to already own the generator at bus $1$. We used hourly bidding and load data from the day-ahead PJM market for 2018~\cite{PJM}. Since load profiles correspond to areas and there are only 21 of them, profiles were randomly assigned to buses. Each bus load profile was perturbed by adding a uniformly distributed deviation of $\pm 5\%$ independently over time and buses. Load profiles were finally scaled so their annual peak matched the benchmark load. For thermal units, we assumed $f_e(\bp_{e,t})=g_e(\bp_{e,t})$, whereas for wind ones we set $f_e(\bp_{e,t})=g_e(\bp_{e,t})=0$. For wind units, we assumed a cost of $3\cdot 10^{6}$~\$/MW for purchase and installation, plus $1\cdot 10^{6}$~\$/MW for operation and maintenance over 25 years. Converting that cost to dollars per hour per unit of active power for a base of $100$~MVA yielded $k=1,826.5$ in \eqref{eq:fun}. All tests were performed on an Intel Core i7 @ 3.4 GHz (16 GB RAM) computer. Problem~\eqref{eq:qp} was solved using the ECOS solver in YALMIP~\cite{ECOS},~\cite{YALMIP}. All times reported are wall-clock times. 


\begin{figure*}[h]
	\centering
	\includegraphics[width=0.25\textwidth]{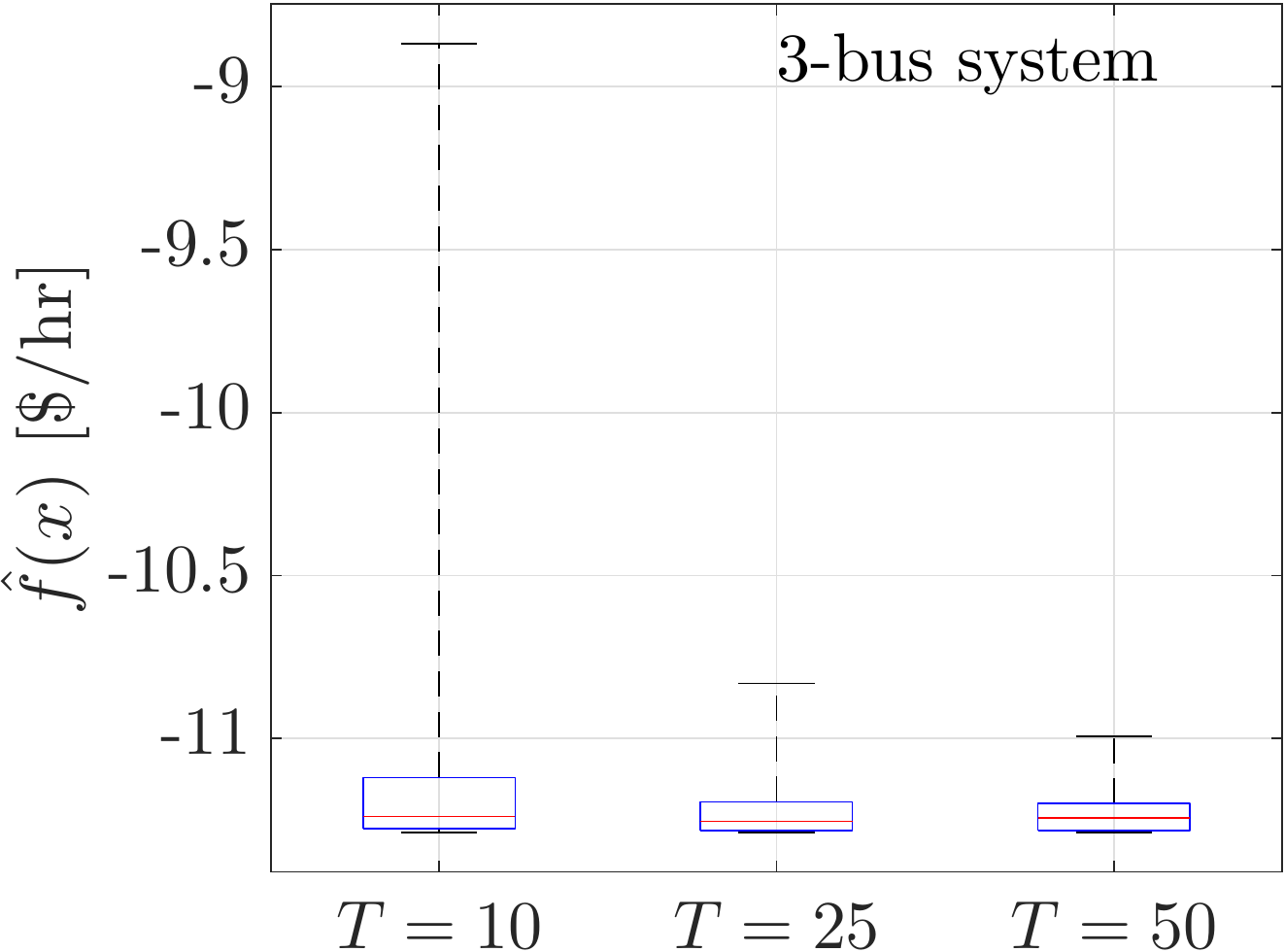}~~~
	\includegraphics[width=0.25\textwidth]{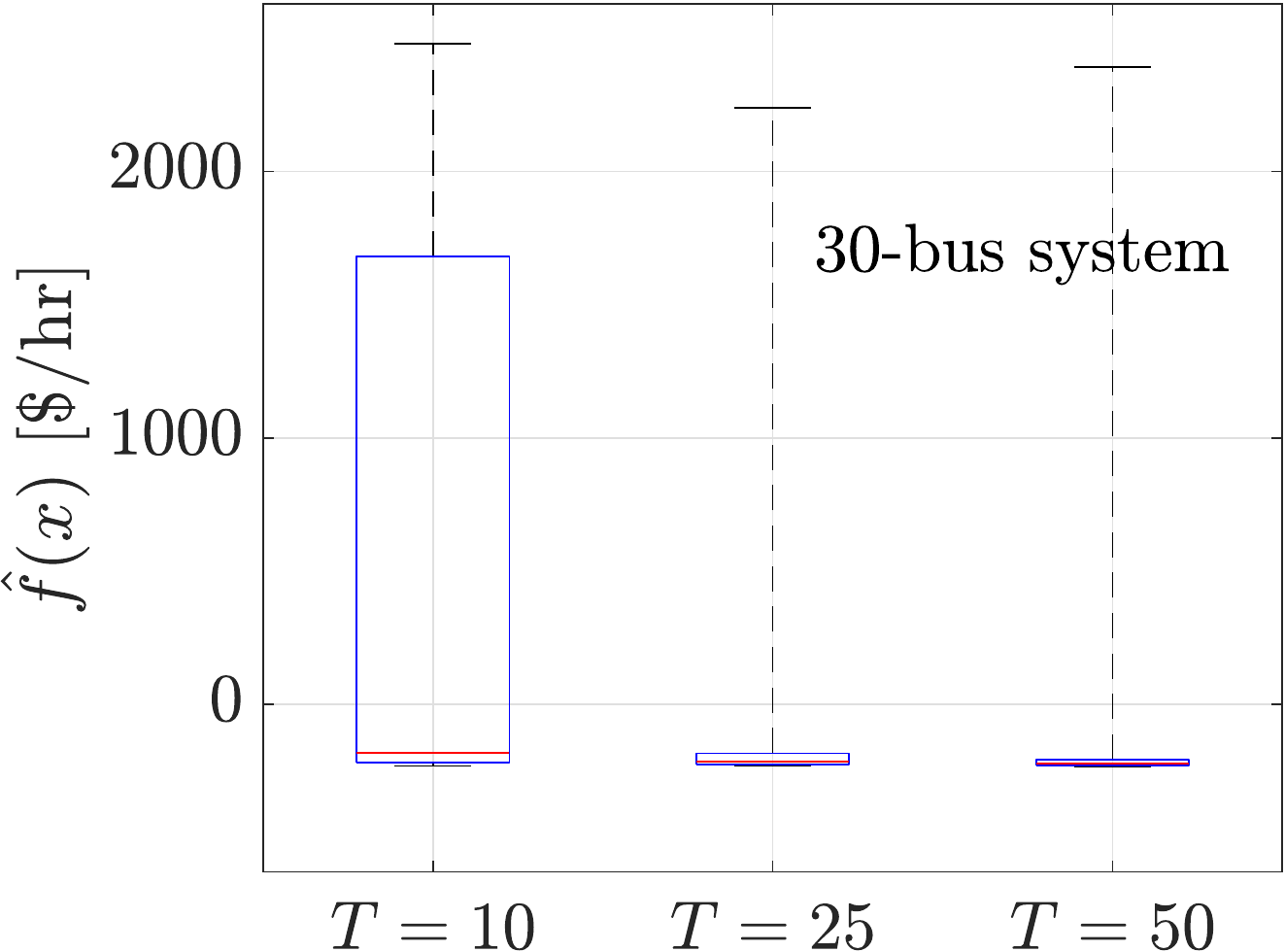}~~~
	\includegraphics[width=0.25\textwidth]{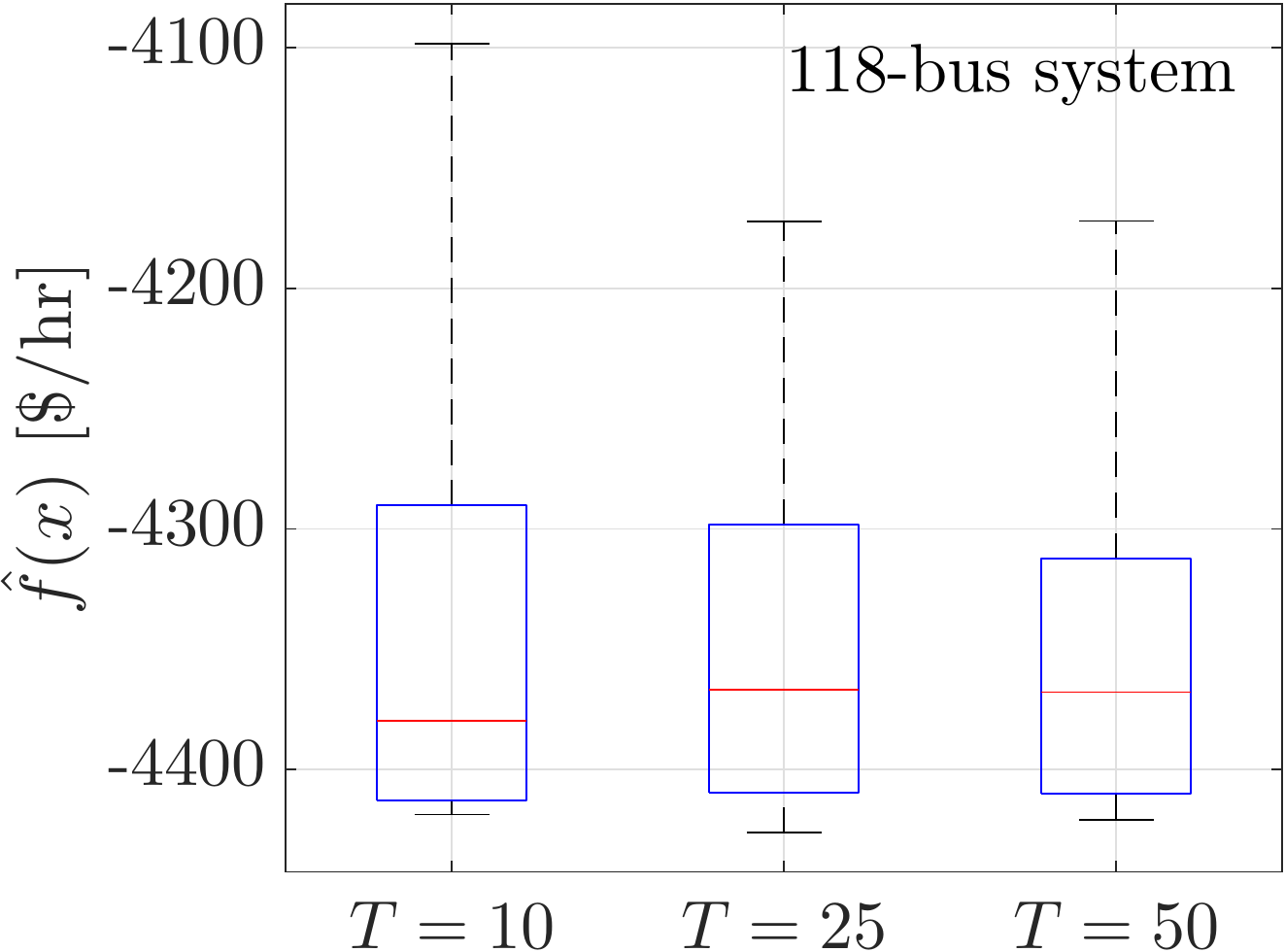}
	\caption{Optimal investment cost attained by the MPEC method of~\cite{KCR12} for the three systems. Although the investment decision $\bx$ has been computed using only $T\in\{10,25,50\}$ scenarios due to computational limitations, the cost shown here is computed over all $8,760$ market scenarios. Box plots are computed over 100 Monte Carlo runs by randomly selecting $T$ scenarios each time. \emph{Left:} For the 3-bus system, the MPEC solutions attained the true optimal cost of $-11.28$ using 25-50 scenarios in 1-5 seconds -- the true optimal cost can be found as the stationary point of $f(x_1)$ in Fig.~\ref{fig:3bus-cost} for $\bar{f}_a$. For the two other systems, the true optimal cost is not known. \emph{Center:} For the 30-bus system, the MPEC solution reaches a seemingly minimal cost using $25$ scenarios, even though the maximum value over the Monte Carlo runs varies widely. \emph{Right:} For the 118-bus system, the sample mean costs lie significantly above the sample minimum ones, and using 50 scenarios is not sufficient to reduce the cost variability.}
	\label{fig:MPEC-cost}
\end{figure*}

\begin{figure}[ht]
	\centering
    \includegraphics[width=0.22\textwidth]{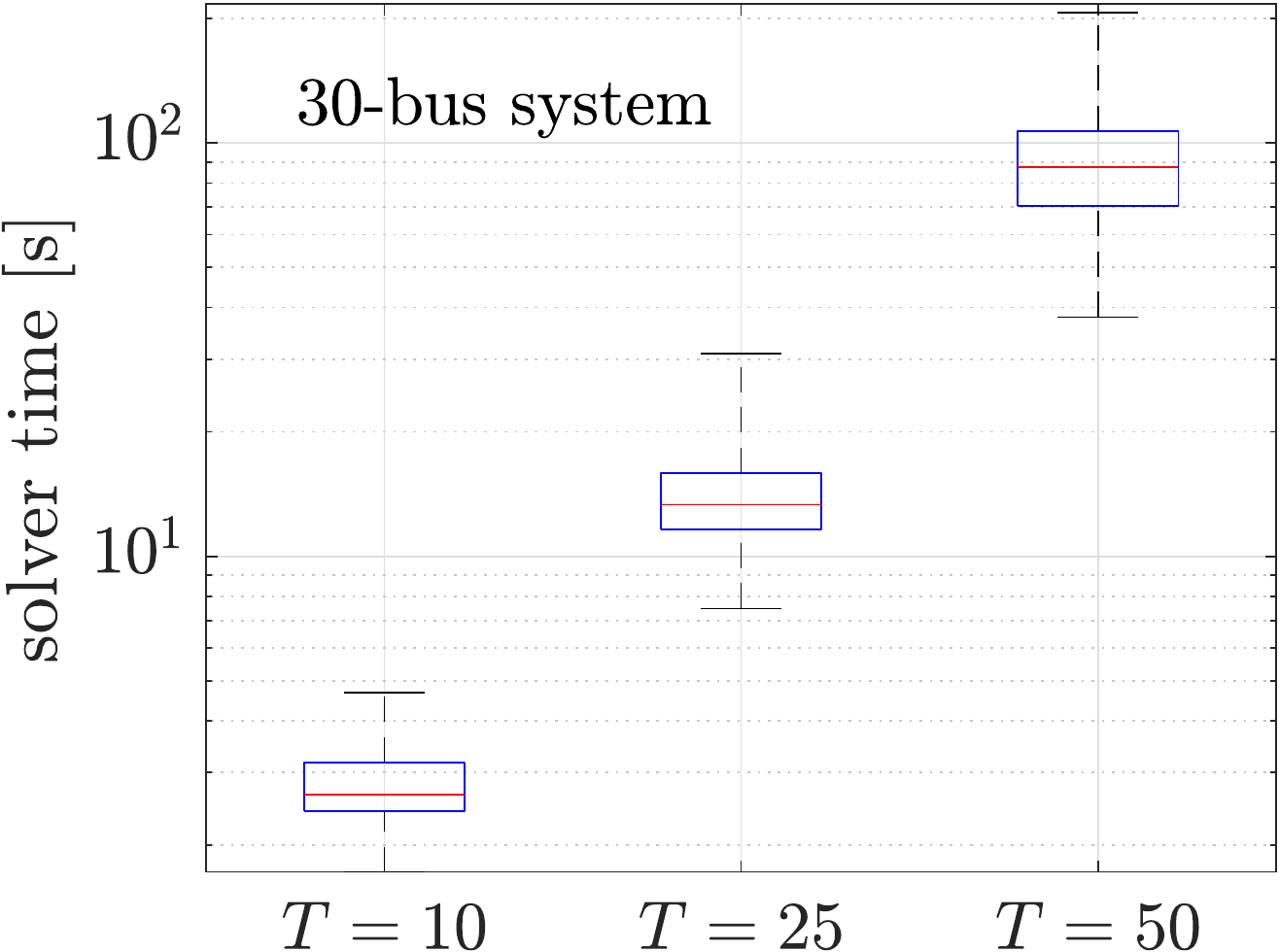}~~
	\includegraphics[width=0.22\textwidth]{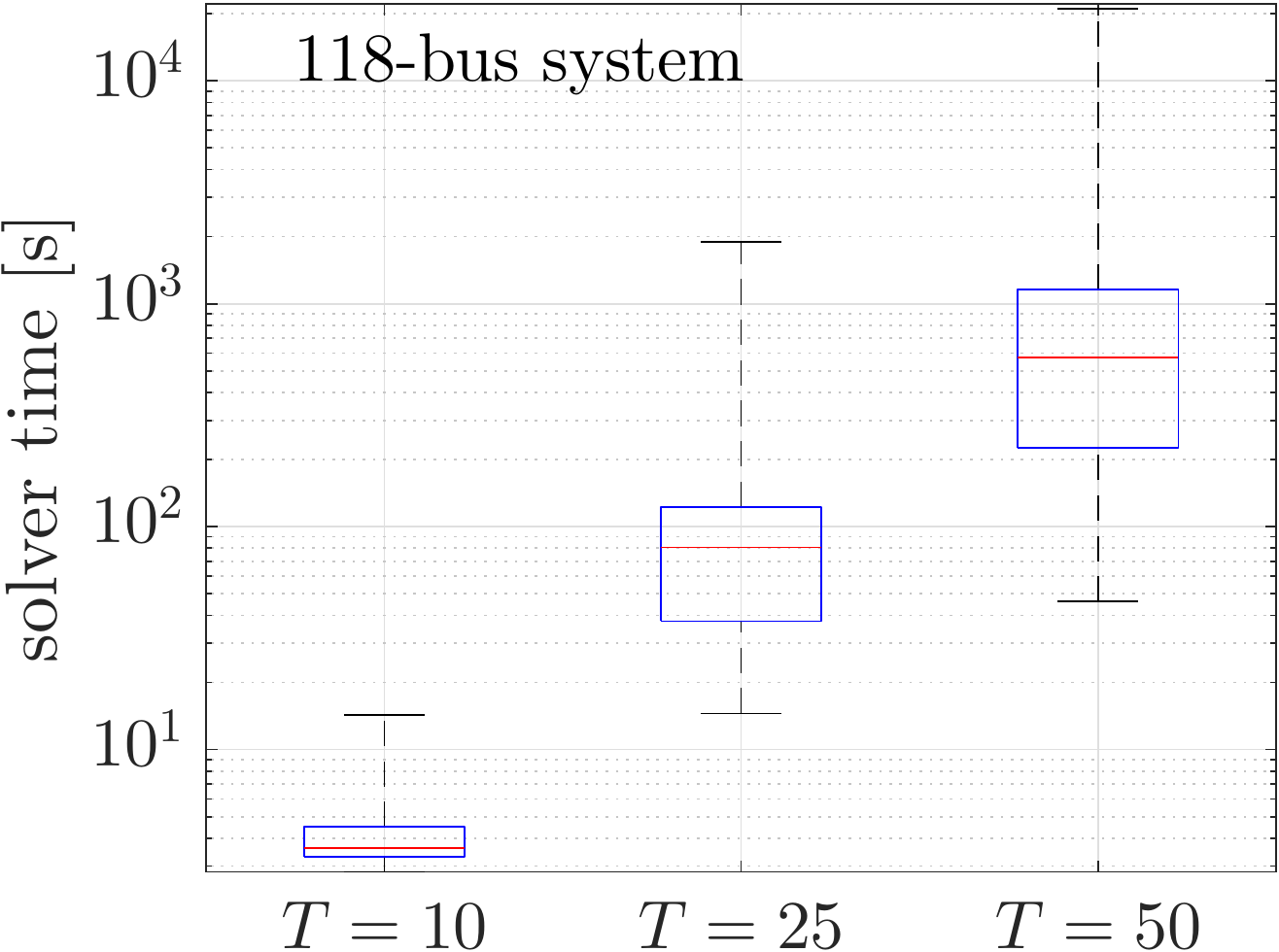}
	\caption{Box plots of running times for the MPEC method of~\cite{KCR12} over 100 Monte Carlo runs by randomly selecting $T$ scenarios each time.}
	\label{fig:MPEC-time}
\end{figure}

The first test explores the effect of the network size and the number of scenarios $T$ on the scalability and optimality of MPEC. MPEC was implemented in Gurobi v.9.0~\cite{gurobi}. The solver's optimality gap was set to $5\%$ and $M=10^4$ in~\eqref{eq:bigm}. Because MPEC could not consider all $8,760$ scenarios, we were able to run MPEC for $T\in\{10,25,50\}$ scenarios per network. Figure~\ref{fig:MPEC-cost} shows the box plots of $\hat{f}(\bx)$ achieved by the MPEC solutions, while Figure~\ref{fig:MPEC-time} shows the box plots for the related running times. For the 3- and 30-bus systems, the MPEC solutions seem to be achieving the optimal cost using only 25-50 scenarios and within reasonable time. For the 118-bus system however, the investment cost varies widely even for 50 scenarios. This observation suggests that finding a meaningful optimizer requires considering $T>50$ scenarios. However, Figure~\ref{fig:MPEC-time} indicates that the running time grows fast with $T$. It is also worth reporting that running MPEC for the 118-bus system and $T=100$ took approximately four days.

\begin{table}[t]
	\renewcommand{\arraystretch}{1.2}
	\caption{Results for Algorithm~\ref{alg:SGS}}
	\label{tbl:GS}
	\centering
	\begin{tabular}{|c|r|r|r|r|}
		\hline\hline
	system & $T\times K$ &\# critical regions & optimal cost & time [s]\\
		\hline\hline
		3-bus&$876,000$&$4$&$-11.31$&$67$\\ 
		\hline
		30-bus&$876,000$&$229$&$-422$&$203$\\
		\hline
		118-bus&$876,000$&$23,695$&$-5,290$&$66,735$
		\\
		\hline\hline
	\end{tabular}
\vspace*{-1.5em}
\end{table}

\begin{table}[t]
	\renewcommand{\arraystretch}{1.2}
	\caption{Optimal Cost Attained by Algorithm~\ref{alg:MPP-SGD} [\$/h]}
	\label{tbl:SGD-cost} \centering
	\begin{tabular}{|c|r|r|r|r|r|}
		\hline\hline
         &\multicolumn{5}{c|}{initialization}\\
        \hline
        system&1&2&3&4&5\\
		\hline\hline
		3-bus&$-11.28$&$-11.28$&$-11.28$&$-11.28$&$-11.28$\\ 
		\hline
		30-bus&$-422$&$-419$&$-417$&$-421$&$-420$\\
		\hline
		118-bus&$-5,280$&$-5,274$&$-5,296$&$-5,256$&$-5,275$
		\\
		\hline\hline
	\end{tabular}
\vspace*{-1.5em}
\end{table}

\begin{table}[t]
	\renewcommand{\arraystretch}{1.2}
	\caption{Running Time of Algorithm~\ref{alg:MPP-SGD} [s]}
	\label{tbl:SGD-time} \centering
	\begin{tabular}{|c|r|r|r|r|r|}
		\hline\hline
         &\multicolumn{5}{c|}{initialization}\\
        \hline
        system&1&2&3&4&5\\
		\hline\hline
		3-bus&$363$&$323$&$260$&$362$&$483$\\ 
		\hline
		30-bus&$2,246$&$4,240$&$4,266$&$2,000$&$2,178$\\
		\hline
		118-bus&$14,100$&$9,747$&$10,743$&$9,495$&$9,507$
		\\
		\hline\hline
	\end{tabular}
\vspace*{-1.0em}
\end{table}

\begin{figure*}[t]
	\centering
	\subfigure{\includegraphics[width=0.27\textwidth]{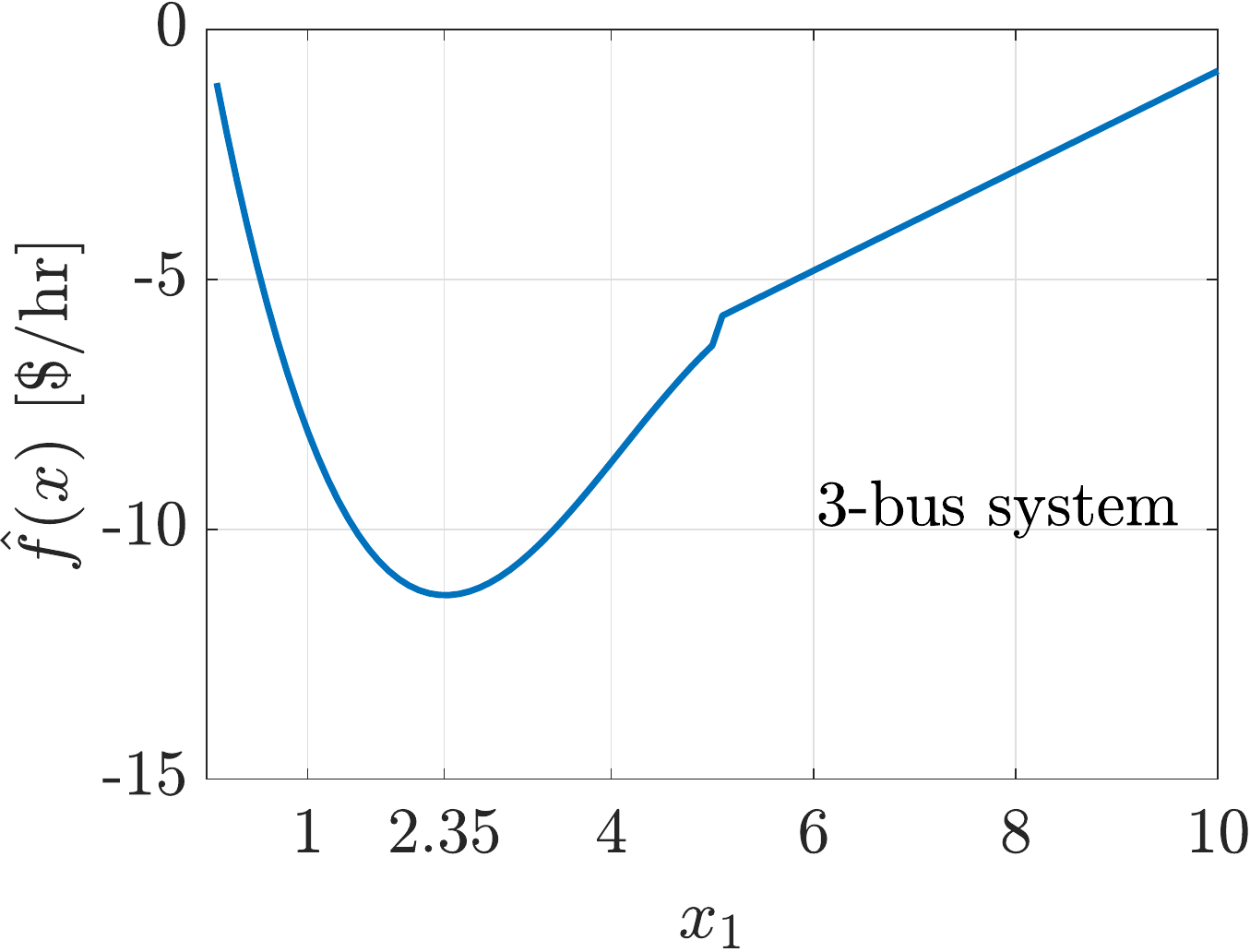}}
	\subfigure{\includegraphics[width=0.27\textwidth]{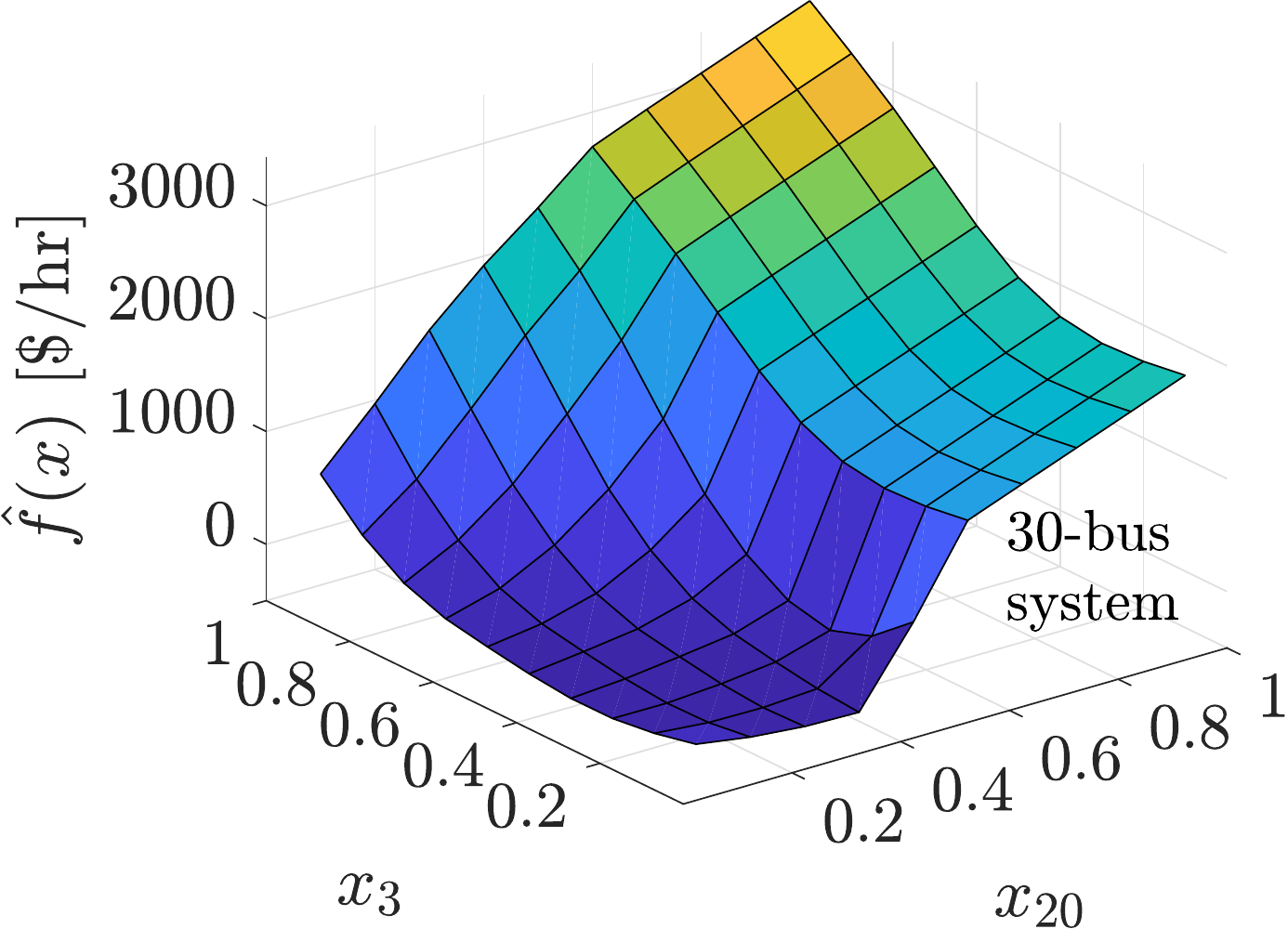}}
	\subfigure{\includegraphics[width=0.30\textwidth]{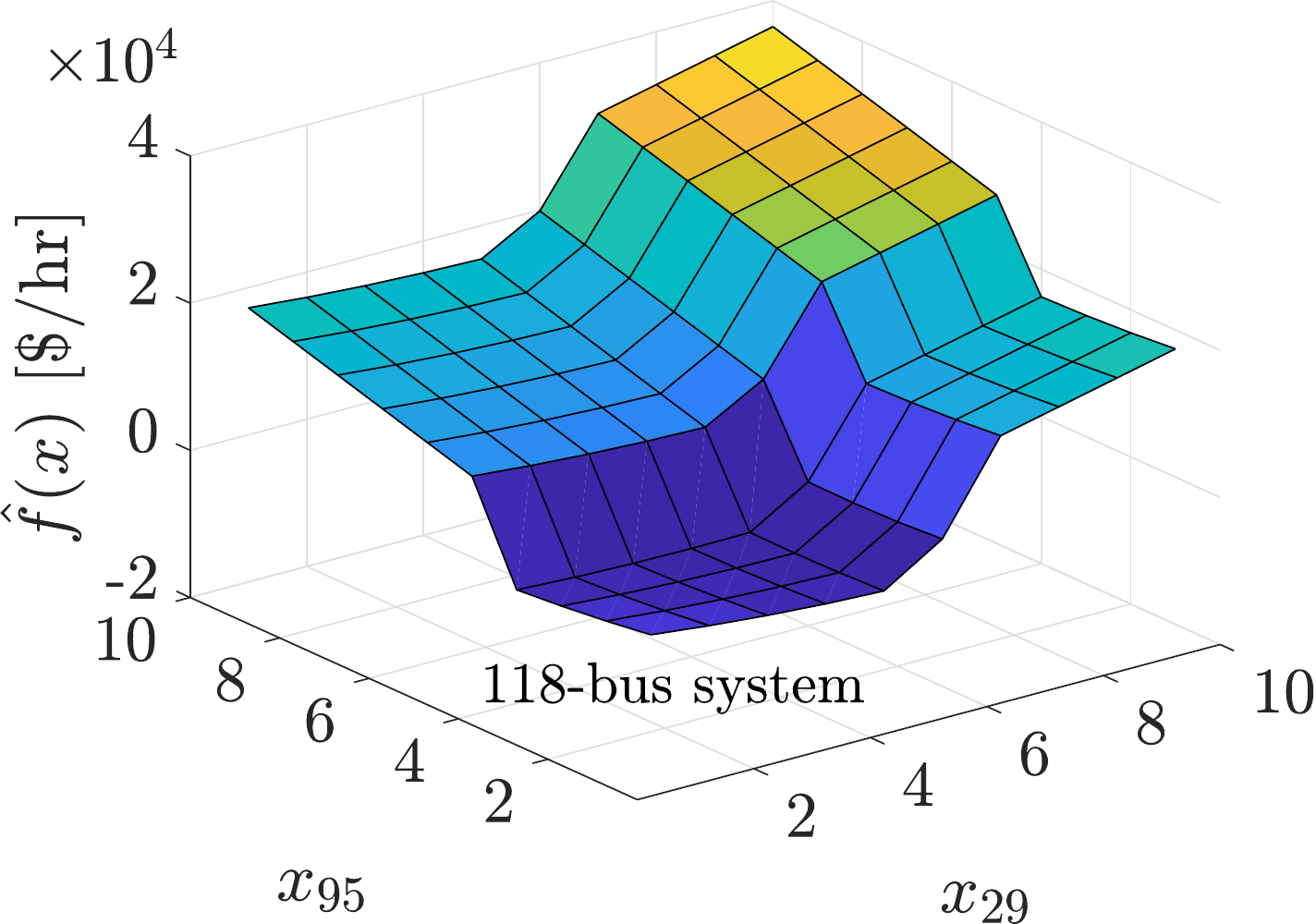}}
	\caption{The value of the investment cost $\hat{f}(\bx)$ for the 3-bus \emph{(left)}; 30-bus \emph{(middle)}; and 118-bus systems \emph{(right)}.}
	\label{fig:GS}
	\vspace*{-.5em}
\end{figure*}

The second test evaluates the optimality and scalability of our Alg.~\ref{alg:SGS} for the MPP-GS method. For the 3-bus system, we considered a 1-D grid of $100$ uniformly-spaced values for $x_1\in\left[0,10\right]$ and $T=8,760$ scenarios drawn from $\ell\in\mathcal{U}(0,10)$. For the 30-bus and 118-bus systems, we considered $10$ uniformly-spaced values for each investment in the range of $\left[0,1\right]$ and $\left[0,10\right]$ for each location, respectively, resulting in a 2-D grid with $100$ points. With $T=8,760$ scenarios, this gave a total of $876,000$ DC-OPFs to be solved as reported in Table~\ref{tbl:GS}. This table also shows the number of critical regions identified; the optimal cost found; and the running times. Figure~\ref{fig:GS} shows the investment cost achieved by Alg.~\ref{alg:SGS}. The MPP-GS was successful in finding an investment cost lower than that of MPEC indicating the advantage of considering the complete scenario set. For the 3- and 30-bus systems, there is also significant computational advantage. The relatively longer time needed for the 118-bus system can be attributed to the large number of critical regions identified that are due to the wider range of $\bx$ and larger system.

\begin{figure*}[t]
	\centering
	\subfigure{\includegraphics[width=0.29\textwidth]{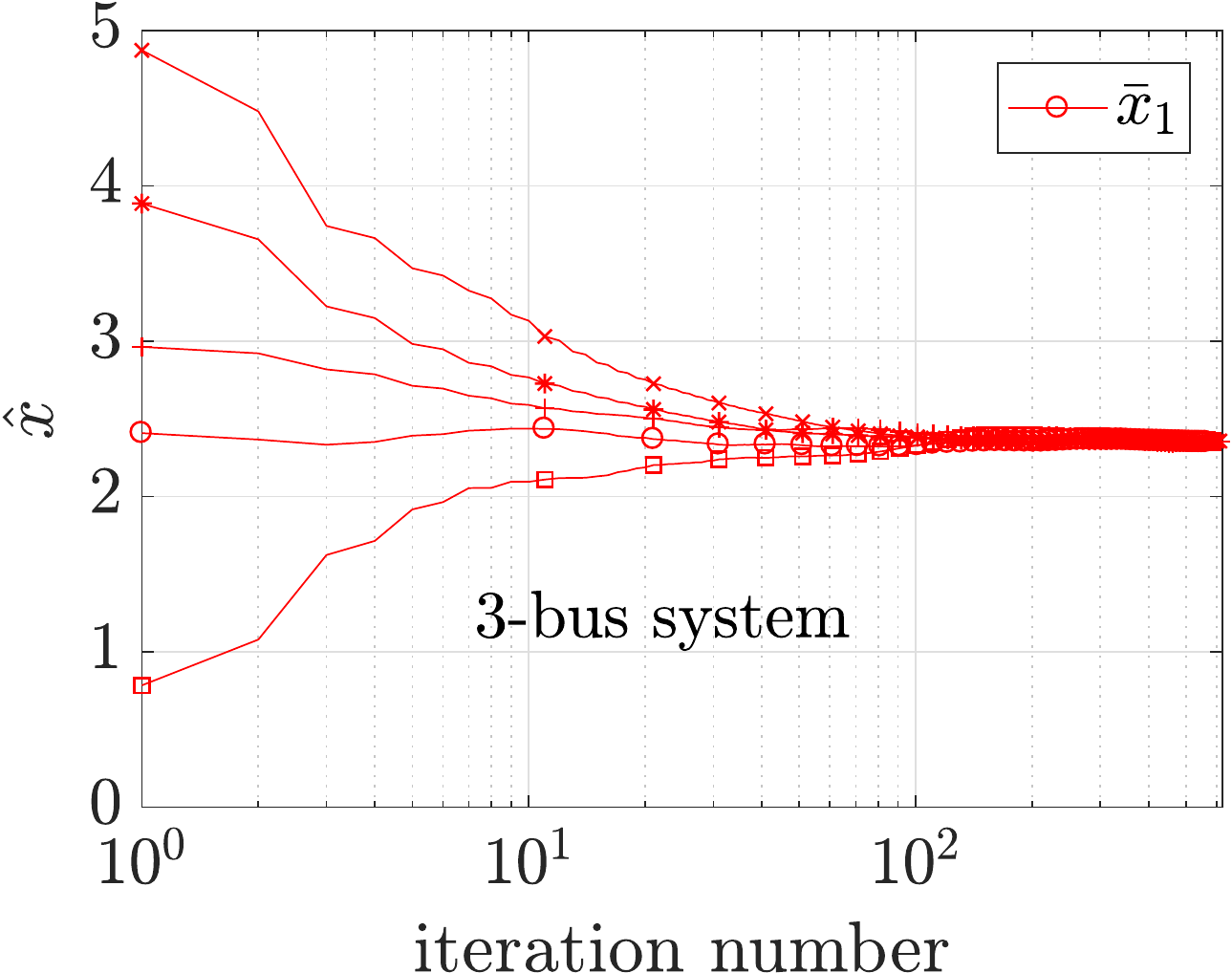}}
	\subfigure{\includegraphics[width=0.32\textwidth]{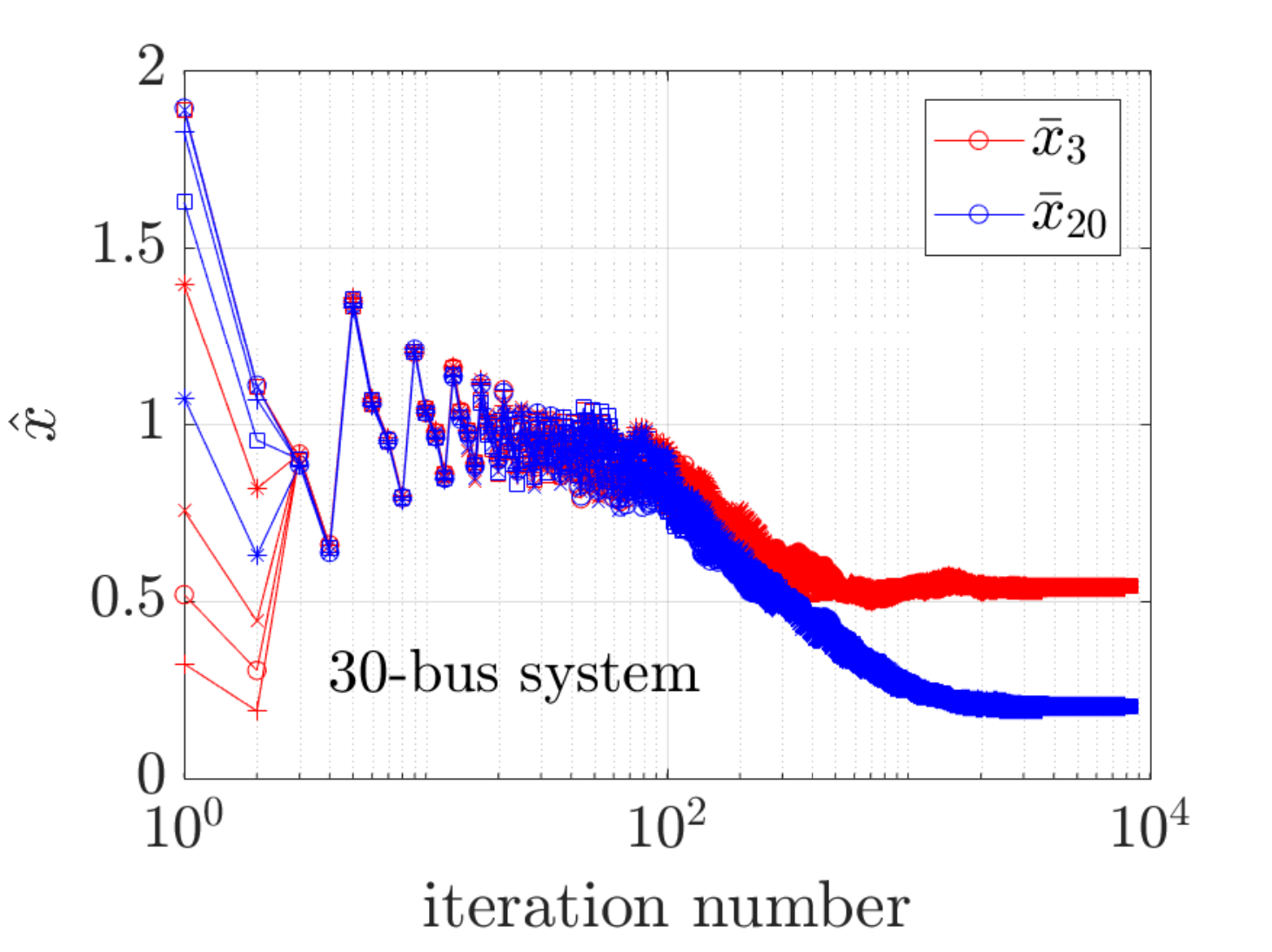}}
	\subfigure{\includegraphics[width=0.32\textwidth]{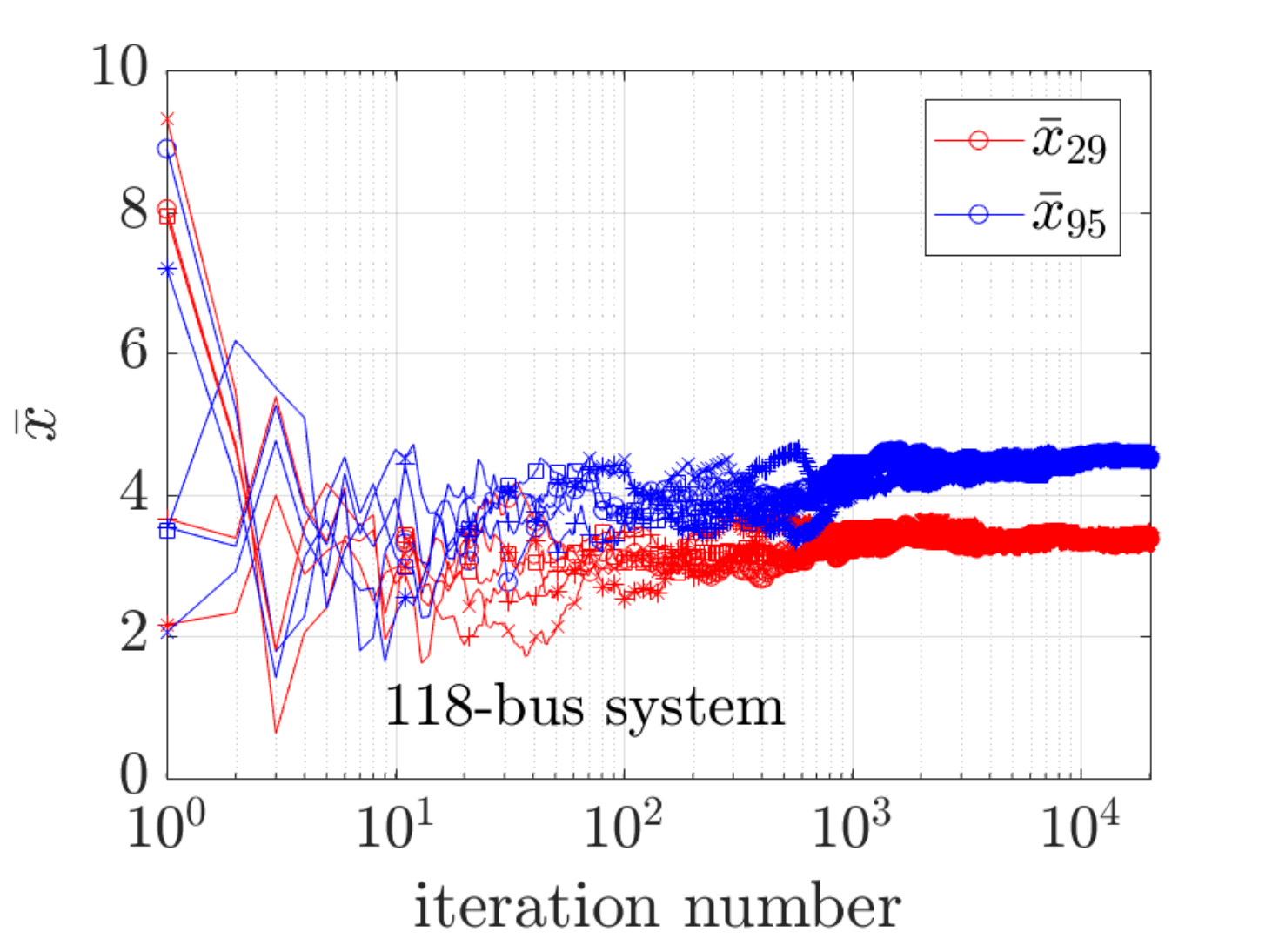}}
 	\vspace*{-1.0em}
	\caption{Convergence of Alg.~\ref{alg:MPP-SGD} for 5 random initializations for the 3-bus system \emph{(left)}; the 30-bus system \emph{(center)}; and the 118-bus system (right). For the 3-bus system, the minimizer is unique and the MPP-SGD algorithm converges to the same point for all initializations.}
	\label{fig:SGD}
	\vspace*{-1.0em}
\end{figure*}

The third test evaluated Alg.~\ref{alg:MPP-SGD} and our MPP-SGD method. Figure~\ref{fig:SGD} shows the convergence of the investment decisions for $5$ randomly initialized trials. Tables~\ref{tbl:SGD-cost} and~\ref{tbl:SGD-time} show the optimal cost and running times, respectively. We observe that for all trials, the optimal cost is much lower than that attained by MPEC for $T=50$ scenarios, which was $-230$ for the 30-bus system and $-4,2208$ for the 118-bus one. Compared to Algorithm~\ref{alg:SGS}, Algorithm~\ref{alg:MPP-SGD} achieves similar optimal costs. The running times of Alg.~\ref{alg:MPP-SGD} are longer for the 3- and 30-bus systems, but much lower for the 118-bus system. Even though the MPP-SGD iterates do not converge to the same decisions for all trials, they attain relatively similar investment costs. This agrees with the findings of Fig.~\ref{fig:GS}, where the cost function seems to be relatively flat at the optimum.

To recapitulate, MPEC can find a globally optimal solution within reasonable time for smaller systems. For larger systems, the complexity involved confines MPEC to relatively few scenarios $T$, which may result in subpar investment solutions if those solutions are to be evaluated on realistic scenario sets. Algorithm~\ref{alg:SGS} is able to achieve much lower average costs over a year-long complete dataset, yet is limited by the number of investment locations and the granularity of the search grid. The latter issues are alleviated by Algorithm~\ref{alg:MPP-SGD}, which seems to be finding near-optimal decisions by handling the complete dataset in running times that improved relatively with the network size.

\section{Conclusions}\label{sec:conclusions}
This work has exploited MPP to devise two SI solvers. The grid search algorithm can handle cases where the number of investment locations is small. Although the needed function evaluations constitute an enormous dataset of DC-OPF instances, their exact primal/dual solutions can be computed upon solving only a limited number of these OPFs, thus accelerating the search by 8-12 times. For larger numbers of investment locations, we have devised a stochastic gradient search scheme, which computes the gradient of the SI objective over entire critical regions in an extremely efficient manner. The developed tools facilitate faster and more educated energy market decisions, while the ideas put forth can be proved fruitful for coping more efficiently with transmission expansion planning and contingency analysis.

\balance
\bibliographystyle{IEEEtran}
\bibliography{myabrv,power}
\end{document}

%% file: stylesV2.tex
\newtheorem{assumption}{Assumption}

\newcommand \bzero{\mathbf{0}}
\newcommand \bone{\mathbf{1}}

\newcommand \bb{\mathbf{b}}
\newcommand \bc{\mathbf{c}}
\newcommand \bd{\mathbf{d}}

\newcommand \bef{\mathbf{f}} 
\newcommand \bg{\mathbf{g}}

\newcommand \bk{\mathbf{k}}
\newcommand \bdell{\boldsymbol{\ell}} 

\newcommand \bp{\mathbf{p}}
\newcommand \bq{\mathbf{q}}
\newcommand \br{\mathbf{r}}

\newcommand \bx{\mathbf{x}}
\newcommand \by{\mathbf{y}}

\newcommand \bA{\mathbf{A}}
\newcommand \bB{\mathbf{B}}
\newcommand \bC{\mathbf{C}}

\newcommand \bE{\mathbf{E}}
\newcommand \bF{\mathbf{F}}

\newcommand \bH{\mathbf{H}}
\newcommand \bI{\mathbf{I}}

\newcommand \bK{\mathbf{K}}

\newcommand \bM{\mathbf{M}}

\newcommand \bQ{\mathbf{Q}}

\newcommand \bS{\mathbf{S}}

\newcommand \balpha{\boldsymbol{\alpha}}

\newcommand \bgamma{\boldsymbol{\gamma}}
\newcommand \bdelta{\boldsymbol{\delta}}

\newcommand \btheta{\boldsymbol{\theta}}

\newcommand \blambda{\boldsymbol{\lambda}}
\newcommand \bmu{\boldsymbol{\mu}}

\newcommand \bpi{\boldsymbol{\pi}}

\newcommand \bomega{\boldsymbol{\omega}}

\newcommand \bDelta{\mathbf{\Delta}}

\newcommand \mcC{\mathcal{C}}

\newcommand \mcP{\mathcal{P}}

\newcommand \mcX{\mathcal{X}}

\newcommand \tbb{\tilde{\mathbf{b}}}

\newcommand \tbA{\tilde{\mathbf{A}}}

\newcommand \tbE{\tilde{\mathbf{E}}}

\newcommand \tblambda{\tilde{\boldsymbol{\lambda}}}

\newcommand \hbp{\hat{\mathbf{p}}}

\newcommand \bbb{\bar{\mathbf{b}}}

\newcommand \bbp{\bar{\mathbf{p}}}

\newcommand \bbx{\bar{\mathbf{x}}}

\newcommand \bbA{\bar{\mathbf{A}}}

\newcommand \bbE{\bar{\mathbf{E}}}

\newcommand \bblambda{\bar{\boldsymbol{\lambda}}}